\documentclass[aps,prx,superscriptaddress,amsmath,amssymb,twocolumn,10pt, epsfig]{revtex4-1}

\pdfoutput=1
\usepackage{appendix}
\usepackage{amssymb, amsmath,amsthm}    
\usepackage{graphicx}   
\usepackage{verbatim}   
\usepackage{color}      
\usepackage{subfigure}  
\usepackage{caption}
\usepackage{tikz}
\usepackage{pgflibraryarrows}
\usepackage{pgflibrarysnakes}
\usepackage{lipsum}
\usepackage{mathtools}
\usepackage{ae} 
\usepackage[T1]{fontenc}
\usepackage[ansinew]{inputenc}
\usepackage[pdfstartview=FitV,linktocpage,colorlinks=true,linkcolor=darkblue,citecolor=darkblue,urlcolor=darkblue]{hyperref}
\usepackage{epsfig}
\usepackage{enumerate}
\usepackage{varioref}
\usepackage{cleveref}
\usepackage{makeidx}
\makeindex
\usepackage[english]{babel}
\usepackage{ulem}

\newcommand{\beq}{\begin{equation}}
\newcommand{\eeq}{\end{equation}}
\newcommand{\beqq}{\begin{equation*}}
\newcommand{\eeqq}{\end{equation*}}
\newcommand\bea{\begin{array}}
\newcommand\eea{\end{array}}
\newcommand\beaa{\begin{array*}}
\newcommand\eeaa{\end{array*}}
\newcommand\beal{\begin{align}}
\newcommand\eeal{\end{align}}
\newcommand\beall{\begin{align*}}
\newcommand\eeall{\end{align*}}

\def\a{{\alpha}}
\def\L{\Lambda}

\def\s{\sigma}

\newcommand{\ep}{\epsilon}
\newcommand{\eps}{\varepsilon}

\def\O{{\mathcal O}}

\newcommand{\f}{\frac}

\newcommand{\Tr}{{\rm Tr}}

\newcommand{\abs}[1]{\lt|{#1}\rt|}

\DeclareMathOperator{\sign}{sgn}

\newcommand{\lt}{\left}
\newcommand{\rt}{\right}
\def\[{\left[}
\def\]{\right]}
\def\({\left(}
\def\){\right)}
\def\<{\langle}
\def\>{\rangle}

\newcommand{\nn}{\nonumber}

\newcommand{\eq}[1]{Eq.(\ref{#1})}

\definecolor{darkblue}{cmyk}{0.9,0.9,0,0}

\begin{document}
	\title{Many-body chaos in the antiferromagnetic quantum critical metal}
	\author{Peter Lunts}
	\affiliation{Center for Computational Quantum Physics, Flatiron Institute, 162 5th Avenue, New York, NY 10010, USA}
	\author{Aavishkar A. Patel}
	\affiliation{Department of Physics, Harvard University, Cambridge MA 02138, USA}
	\date{\today}
	
\begin{abstract}
We compute the scrambling rate at the antiferromagnetic (AFM) quantum critical point, using the fixed point theory of \href{https://journals.aps.org/prx/abstract/10.1103/PhysRevX.7.021010}{Phys. Rev. X $\boldsymbol{7}$, 021010 (2017)}. At this strongly coupled fixed point, there is an emergent control parameter $w \ll 1$ that is a ratio of natural parameters of the theory. The strong coupling is unequally felt by the two degrees of freedom: the bosonic AFM collective mode is heavily dressed by interactions with the electrons, while the electron is only marginally renormalized. We find that the scrambling rates act as a measure of the ``degree of integrability'' of each sector of the theory: the Lyapunov exponent for the boson $\lambda_L^{(B)} \sim \O(\sqrt{w}) \,k_B T/\hbar$ is significantly larger than the fermion one $\lambda_L^{(F)} \sim \O(w^2) \,k_B T/\hbar$, where $T$ is the temperature. Although the interaction strength in the theory is of order unity, the larger Lyapunov exponent is still parametrically smaller than the universal upper bound of $\lambda_L=2\pi k_B T/\hbar$. We also compute the spatial spread of chaos by the boson operator, whose low-energy propagator is highly non-local. We find that this non-locality leads to a scrambled region that grows exponentially fast, giving an infinite ``butterfly velocity'' of the chaos front, a result that has also been found in lattice models with long-range interactions. 

\end{abstract}

\maketitle

\section{Introduction}

Thermalization in closed quantum systems \cite{PhysRevA.43.2046, PhysRevE.50.888, PhysRevLett.80.1373, Rigol_Nature} is an important topic in quantum dynamics that has recently received much attention \cite{Langen_Nature_physics, Jurcevic_Nature, Kaufman794, Schreiber842, Choi1547, PhysRevLett.114.083002, Meldgin_Nature, Smith_Nature}. Thermalizing quantum many-body systems are ``non-integrable", i.e. they do not have a number of conserved quantities that is of the order of the large number of degrees of freedom in the system. 
\\
\indent 
A candidate quantitative measure of physical processes that destroy integrability and lead to thermalization is the degree of ``many-body chaos" or ``scrambling" as computed from certain special correlators \cite{1969JETP...28.1200L, Shenker2014, Kitaev1}. These correlators measure the growth of the commutator of two local operators in time. Given operators $A,B$ of a quantum many-body system with a Hamiltonian $H$ at inverse temperature $\beta$, the correlator at time $t$ is defined by
\begin{equation}
f_{A,B}(t) = \Tr\[e^{- \beta H/2} \[A(t),B\] e^{- \beta H/2} \[A(t),B\]^{\dagger} \].
\label{eq:f definition}
\end{equation}
As opposed to the usual thermal trace, the thermal density matrix has been split up into two parts and inserted between the two commutators in order to avoid short-distance divergences of the continuum field theory, thereby ``regularizing" the correlator. The operators $A$ and $B$ are chosen such that $f_{A,B}(t)$ starts out very small at $t=0$. At late times, $f_{A,B}(t)$ will saturate to an $\O(1)$ value. At intermediate times, for non-integrable systems, the out-of-time-ordered (OTO) terms in $f_{A,B}(t)$ are expected to lead to a rapid growth $f_{A,B}(t) \sim \ep \, e^{\lambda_L^{(A,B)} \, t}$ \cite{Kitaev1}. The exponent $\lambda_L^{(A,B)}$ is called a ``Lyapunov exponent" and generically quantifies the non-integrability of the quantum many-body system \footnote{There have been recent works showing that there are some exceptions to this \cite{PhysRevB.98.060302, PhysRevB.98.220303}}. The small parameter $\ep\ll 1$ enables the exponential growth to be defined as $f_{A,B}(t)$ grows from an $\mathcal{O}(\ep)$ value at $t=0$ to an $\mathcal{O}(1)$ value at later times. Certain many-body quantum systems with a large number, $N$, of local degrees of freedom have $\epsilon\sim\mathcal{O}(1/N)$ even if $A,B$ are spatially close to each other. However, in systems with an $\O(1)$ number of local degrees of freedom, a large spatial separation between local operators $A,B$ is required to have a small $\epsilon$ \cite{Kitaev1, Stanford2016, Patel21022017, PhysRevD.96.065005, 2017arXiv170507895W, PhysRevB.98.045102, 2018arXiv180512299J,PhysRevX.7.031047,ALEINER2016378, PhysRevB.95.134302,PhysRevB.98.144304,2018arXiv180200801X}.  
\\
\indent 
The largest Lyapunov exponents are expected to occur in many-body quantum systems without quasiparticles, since the complete lack of particle coherence in any basis of operators implies the non-existence of a thermodynamic number of conserved quantities and strong non-integrability. The upper bound of $\lambda_L = 2 \pi k_B T/\hbar$ (from here on we set $k_B = \hbar = 1$) \cite{Maldacena2016} is saturated by black holes \cite{Shenker2014} and certain Sachdev-Ye-Kitaev models of strange metals \cite{PhysRevLett.70.3339, Kitaev1}. On the other hand, Fermi liquids, for example, have $\lambda_L \sim T^2$ \cite{ALEINER2016378, PhysRevB.95.134302}. Several works so far have computed $\lambda_L$ for various weakly and strongly coupled systems \cite{Kitaev1, Stanford2016, Patel21022017, PhysRevD.96.065005, 2017arXiv170507895W, PhysRevB.98.045102, 2018arXiv180512299J,PhysRevX.7.031047,ALEINER2016378, PhysRevB.95.134302}. However, $\lambda_L$ has not yet been computed analytically in a strongly coupled system without a large number of flavors.
\\
\indent 
In this work we study many-body quantum chaos at the antiferromagnetic (AFM) quantum critical point (QCP) of a two-dimensional metal, believed to exist in a wide range of layered compounds \cite{PhysRevLett.105.247002, Hashimoto22062012, park2006hidden}. It is described by the spin-fermion model \cite{PhysRevLett.84.5608, doi:10.1080/0001873021000057123, PhysRevLett.93.255702, PhysRevB.82.075128, Abrahams28022012, PhysRevB.91.125136, PhysRevB.93.165114, Berg21122012, PhysRevLett.117.097002, 2016arXiv160909568W, PhysRevB.95.245109, PhysRevB.98.075140, PhysRevX.7.021010}, in which the fluctuations of the AFM collective bosonic mode scatter electrons between pairs of ``hot spots''. The stable low energy fixed point of this model was found in Ref. \cite{PhysRevX.7.021010}. This strongly coupled fixed point occurs near perfect nesting of the paired hot spots, and deviations from it are perturbative in an emergent dynamical control parameter, $w$, which is a ratio of natural parameters of the theory. The electrons are perturbed away from free particles only by marginal corrections of strength $w \, \log(\L/T)$ (where $\L$ is a UV cutoff), whereas the bosonic collective excitation is made incoherent and non-local by strong renormalization from the electrons. 
\\
\indent 
We compute two Lyapunov exponents at this fixed point, for the scrambling of both the fermion and boson operators. We do this perturbatively in $w$, without a large number of flavors. Since the interaction strength is $\O(1)$, one might expect the two scrambling rates to be the same and that both scale as $\lambda_L \sim \O(1) \, T$. However, the strong coupling is felt disproportionately by the boson, as the dressing of the fermion is of the order of the small parameter $w$. We find that the fermion exponent $\lambda_L^{(F)} \sim \O(w^2) \, T$ is much smaller than the boson exponent $\lambda_L^{(B)} \sim \O(\sqrt{w}) \, T$ (both relations are up to logarithms in $w$ and $T$). Their large difference is a reflection of the fact that the degrees of renormalization of the two sectors are wildly different, even though at low energies they interact strongly. In this sense, the Lyapunov exponents here act as measures of the ``degrees of integrability'' of the two sectors of the theory. However, contrary to the intuition about systems with strong interactions, the larger exponent  $\lambda_L^{(B)}$ is parametrically smaller than $\O(1) \, T$. This is therefore a critical point where the effective coupling is strong ($\O(1)$), and the boson is heavily renormalized at low energies, but the degree of non-integrability is determined by a parameter other than the effective coupling, and the Lyapunov exponents are not close to the upper bound.  
\\
\indent 
Since the boson scrambles much more strongly, we additionally compute the spatial structure of its scrambling, by separating out the two boson operators in Eq. (\ref{eq:f definition}) at large distances. We find that because of the non-local nature of the boson propagator, boson operators spread exponentially fast through the system. In particular, the front of this ``chaotic wave'' is not a linear light-cone, and a finite ``butterfly velocity'' cannot be defined. This is similar to systems with non-local interactions, where Lieb-Robinson bounds \cite{LiebRobinson} break down \cite{Hastings2006, PhysRevLett.113.030602, PhysRevLett.114.157201}. 


\section{Model}

\begin{figure}[!ht]
	\centering
	\subfigure[]
	{\includegraphics[scale=1.]{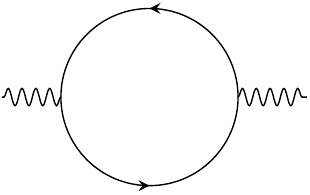}
		\label{fig:SEb}} ~~~~~
	\subfigure[]
	{\includegraphics[scale=1.]{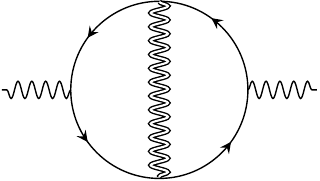}
		\label{fig:2LSEb}} ~~~~~
	\caption{
		The leading order diagrams contributing to the boson self-energy
		in the small $v$ limit.   
		Solid lines are the bare fermion propagators.
		The wiggly double line is the boson propagator 
		consistently dressed with the self-energy in (a) and (b).
		The RPA correction in (a) gives the leading order frequency contribution, while the diagram in (b) gives the leading order momentum contribution. 
	}
	\label{fig: CT1}
\end{figure}

\begin{figure}[h]
	\includegraphics[width=3.5in]{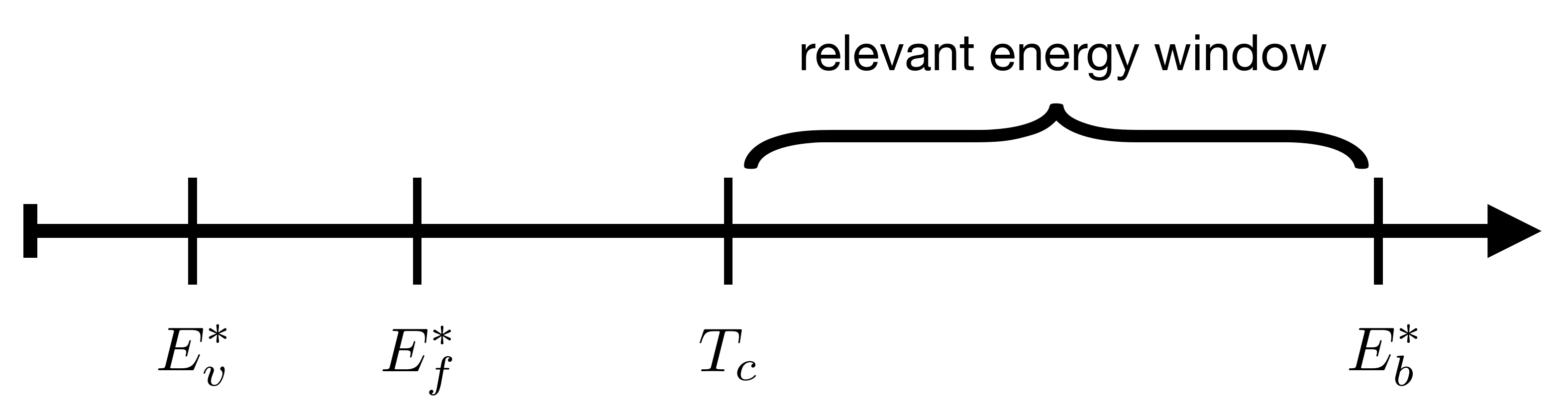}
	\caption{The energy scales associated with the action of Eq. (\ref{eq:2D_theory}). The largest energy scale is $E_b^* \sim c(v)^2 \tilde \Lambda$, where $\tilde \Lambda$ is the UV scale associated with the irrelevant bare kinetic term for the boson. Below $E_b^*$ the boson dynamics is given by the non-local form, and $E_b^*$ acts as a frequency cutoff for the theory of Eq. (\ref{eq:2D_theory}). The second scale down is the superconducting transition temperature, $T_c \sim c(v) \sqrt{\L \tilde \L} \, e^{-\sqrt{c(v)/(\gamma v)}}$, where $\gamma \sim 1$ and $\Lambda$ is the momentum cutoff for the theory of Eq. (\ref{eq:2D_theory}). The third scale down is $E_f^* \sim \L e^{-(\pi/3) \sqrt{(\log 1/v)/v}}$, below which the fermions lose coherence. The smallest scale is $E_v^* \sim \L e^{-1/(v \log 1/v)}$, below which $v$ starts to deviate appreciably from $v_0$. The large energy window relevant to this paper is $T_c < E < E_b^*$.}
	\label{fig:energy scales}
\end{figure}

The spin-fermion model \cite{PhysRevLett.84.5608, doi:10.1080/0001873021000057123, PhysRevLett.93.255702, PhysRevB.82.075128, Abrahams28022012, PhysRevB.91.125136, PhysRevB.93.165114, Berg21122012, PhysRevLett.117.097002, 2016arXiv160909568W, PhysRevB.95.245109, PhysRevB.98.075140, PhysRevX.7.021010} describes a system of two-dimensional electrons interacting with a collective spin excitation at a finite ordering wavevector. This interaction is strongest at certain ``hot spots'' on the Fermi surface that are connected by the ordering wavevector. For simplicity, we take the case of a square lattice with one band, and the wavevector to be $(\pi,\pi)$ (this is most similar to the situation in the cuprates). The Gaussian tree level action has been extensively studied, so instead we start by writing the effective Euclidian action for the low-energy fixed point of this model found in Ref. \cite{PhysRevX.7.021010} at finite temperature,
\begin{widetext}
\begin{equation}
\begin{split}
\mathcal{S} = & 
\sum_{n=1}^4 \sum_{m=\pm} \sum_{\sigma=\uparrow,\downarrow} 
T \sum_{\omega_k}
\int \frac{d\vec k}{(2\pi)^2} \,
{\psi}^{(m)*}_{n,\s}(k)
\left[ i \omega_k + e^{m}_n(\vec k;v)  \right] 
\psi^{(m)}_{n,\sigma}(k) 
\\ &
\hspace{-5mm} + \f12 T \sum_{\omega_q} \int \frac{d\vec q}{(2\pi)^2} \,
\Bigl[
\abs{\omega_q} + c(v) (\abs{q_x} + \abs{q_y}) + M(T,\L,v)
\Bigr]
\vec \phi(q) \cdot \vec \phi(-q)
\\ & 
+ \sqrt{ \frac{\pi v}{2} }  \sum_{n=1}^4 
\sum_{\s,\s'=\uparrow,\downarrow}
T \sum_{\omega_k}
\int \frac{d\vec k}{(2\pi)^2}  
\, T \sum_{\omega_q}
\int \frac{d\vec q}{(2\pi)^2}   
\Bigl[
\vec{\phi}(q) \cdot 
{\psi}^{(+)*}_{n,\s}(k+q) \vec{\tau}_{\s,\s'}  
\psi^{(-)}_{n,\s'} (k) 
+ c.c. \Bigr]. 
\end{split}
\label{eq:2D_theory}
\end{equation}
\end{widetext}
\noindent
Here, $k = (\omega_k, \vec{k})$ denotes the fermionic Matsubara frequency and the two-dimensional momentum $\vec{k} = (k_x, k_y)$. $\psi_{n,\s}^{(m)}$ are the fermion fields that carry spin $\s = \uparrow, \downarrow$ at the hot spots labeled by $n=1,2,3,4, ~ m = \pm$. 
\begin{figure}[h]
	\includegraphics[width=2.5in]{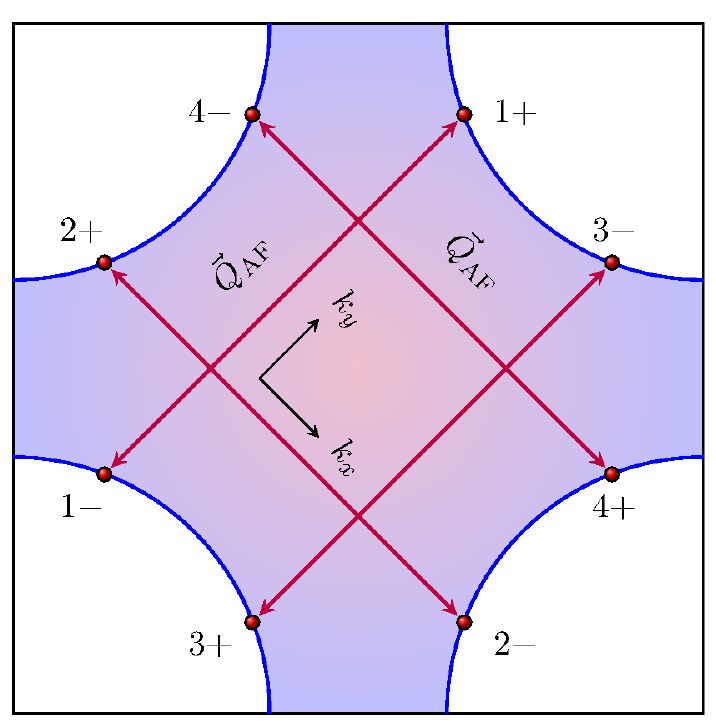}
	\caption{
		The first Brillouin zone of a metal in two dimensions  with $C_4$ symmetry.
		The occupied states live in the shaded region. 
		The AFM ordering wavevector $\vec{Q}_{AFM}$ is denoted by red arrows.
		The hot spots are the red dots connected by $\vec{Q}_{AFM}$. In this minimal model for a generic filling there are eight hot spots.
	}
	\label{fig:hot spots}
\end{figure}
With this choice of axis the ordering wave vector is $\vec{Q}_{AFM} = \pm \sqrt{2} \pi \hat{k}_x, \pm \sqrt{2} \pi \hat{k}_y$ up to the reciprocal lattice vectors $\sqrt{2} \pi (\hat{k}_x \pm \hat{k}_y)$. See Fig. \ref{fig:hot spots} for details. The fermion dispersions are given up to linear order by
$e^{\pm}_1(\vec k;v) = -e^{\pm}_3(\vec k;v) = v k_x \pm k_y$, 
$e^{\pm}_2(\vec k;v) = -e^{\pm}_4(\vec k;v) = \mp k_x + v k_y$, 
where $\vec{k}$ is the momentum deviation from each hot spot. 
The curvature of the Fermi surface can be ignored, since the patches of Fermi surface connected by the ordering vector are not parallel to each other for $v \neq 0$, i.e. the problem is fully two-dimensional. The component of the Fermi velocity along the ordering vector has been set to unity by rescaling $\vec{k}$. 
$v$ is the component of Fermi velocity that is perpendicular to $\vec{Q}_{AFM}$. It controls the degree of nesting between coupled hot spots. A necessary criterion for the validity of Eq. (\ref{eq:2D_theory}) is that $v \ll 1$, i.e. the fermions are close to perfect nesting. This is because, as explained below, a power of $v$ acts as a control parameter for the theory. $\vec{\phi}(q)$ is the boson field with three components
which describes the  AFM collective mode 
with frequency $\omega_q$ and momentum $\vec{Q}_{AFM} + \vec{q}$. 
$\vec{\tau}$ represents the three generators of the $SU(2)$ group. Due to the irrelevance of all local (in space and time) boson terms, there is a freedom to re-scale the boson field (the coefficients of the non-local terms are generated from the fermions and are not independent parameters). This freedom is used to set the Yukawa coupling between the collective mode and the electrons to $\sqrt{\pi v/2}$. The non-local dynamics of the boson is generated from the leading order contributions to the Schwinger-Dyson equation, which are shown in Fig. \ref{fig: CT1}.
The `velocity' $c(v)$ of the strongly damped boson is given by
\begin{equation}
c(v) = \frac{1}{4} \sqrt{  v \log (1/v) },
\label{eq:cv}
\end{equation}
to the leading order in $v$. $M(T,\L,v)$ is the thermal mass with $\Lambda$ being the momentum cutoff in Eq. (\ref{eq:2D_theory}). The leading order in $v$ non-zero contribution to $M(T,\L,v)$ is computed in Appendix \ref{sec:thermal mass}.
\\ \indent 
Although the Yukawa coupling, originally denoted as $g$, is scaled to be $\propto \sqrt{v} \ll 1$, the theory is strongly coupled. The actual strength of interactions, or effective coupling, is given by $g^2/v$ \cite{PhysRevLett.84.5608}, which after our rescaling is $\O(1)$. The strength of interactions is what determines the prefactor of the leading kinetic term for the boson, which comes from the RPA correction and is given by $\abs{\omega_q}$. By choosing this scaling, the boson kinetic term is of the same order as the fermion one, $i \omega_k$. Despite this, the leading order momentum dependence of the boson comes with a small prefactor of $c(v)\sim \sqrt v$. This dependence states that at this fixed point the boson motion is entirely due to the motion of fermions in the direction perpendicular to $\vec Q_{AFM}$. Of course, the generated momentum dependence is still larger than that of the bare kinetic term, $c_0^2 \abs{\vec q}^2/\tilde \Lambda$, below momenta of order $q \sim c(v) \tilde \L/c_0$, where $c_0$ is the bare velocity and $\tilde \Lambda$ is a large UV scale. 
\\ \indent 
The action in Eq. (\ref{eq:2D_theory}) obeys the $z=1$ critical scaling.
All operators are marginal (this is called the interaction-driven scaling \cite{PhysRevB.90.045121, PhysRevB.98.075140}). It turns out that all quantum corrections to Eq. (\ref{eq:2D_theory}) are controlled by 
\begin{equation}
w(v) \equiv \frac{v}{c(v)} \ll 1.
\label{eq:w}
\end{equation}
A renormalization group analysis shows that $v$ flows to zero with decreasing energy scale \footnote{A similar ratio of velocities was found to act as a control parameter for the low energy fixed point of the theory of nematic ordering in $d$-wave superconductors \cite{PhysRevB.78.064512}}. Therefore, there exists a basin of attraction around $v=0$ of finite size where this low energy fixed point is stable. In our work we assume a bare value of $v$ within this basin of attraction and small enough to satisfy Eq. (\ref{eq:w}).
\\
\indent 
There are several points to make here regarding the energy scales present. First, since the interaction in Eq. (\ref{eq:2D_theory}) is marginal, the perturbative in $w(v)$ renormalization of the fermion propagator leads to marginal Fermi liquid behavior at energies/temperatures low enough for quantum corrections to become important. Therefore, the renormalizations of the collective mode and fermion are highly asymmetric: the fermion renormalizes the collective mode to a highly incoherent and non-local form at relatively high energies, i.e. below the cutoff for the theory in Eq. (\ref{eq:2D_theory}), while the feedback onto the fermion remains weak down to much lower energies. In between these two scales is a superconducting transition temperature. Finally, the flow of $v(\mu) \rightarrow 0$ with decreasing energy scale $\mu$ will introduce an additional energy/temperature dependence into the propagators, beyond the usual logarithmic renormalization corrections. However, this flow is logarithmic in $\mu$, and therefore in order to see $v$ change significantly from its bare value $v_0$ the energy scale $\mu$ must be extremely small, in particular much less than the scale at which the fermions lose coherence. These various energy scales are shown in Fig. \ref{fig:energy scales}. 
We are interested in the large energy window above the superconducting transition temperature and below the cutoff for the fixed point theory. Therefore, in particular, we can ignore the flow of $v(\mu)$ and treat it as $v = v_0$.

\section{Computing many-body chaos}

We are interested in computing the many-body chaos (or ``scrambling") in both the fermion and boson sectors of the theory. This requires the computation of two correlators of the type in Eq. (\ref{eq:f definition}): one with fermion and one with boson creation/annihilation operators. They are given by
\begin{align}
\begin{split}
f_F(t,\vec x) &= \theta(t) 
\frac{1}{2^2}
\sum_{\sigma, \sigma'=\uparrow,\downarrow}
\Tr \lt[ e^{-\beta H/2} \, \{{\psi}_{\s} (\vec x,t), {\psi}^{*}_{\s'} (0) \} 
\rt.\\ & \lt.
\hspace{30mm} 
e^{-\beta H/2} \, \{{\psi}_{\s} (\vec x,t), {\psi}^{*}_{\s'} (0) \}^{\dagger} \rt],
\end{split}
\label{eq:f_F definition}
\\
\begin{split}
f_B(t, \vec x) &= \theta(t) \frac{1}{3^2} \sum_{i,j = 1}^{3} \Tr \lt[ e^{-\beta H/2} \, [\phi_i (\vec x,t), \phi_j (0)] 
\rt.\\ & \lt.
\hspace{30mm} 
e^{-\beta H/2} \, [\phi_i (\vec x,t), \phi_j (0)]^{\dagger} \rt]. 
\end{split}
\label{eq:f_B definition}
\end{align}
These correlation functions are generated from an action defined on a complex-time contour that has two real-time folds separated by $i \beta/2$. We compute $f(t,\vec x)$ perturbatively in $w(v)$ by solving a Bethe-Salpeter equation for the Fourier transform $f(\omega, \vec p)$ that gives a resummation of an infinite number of ladder diagrams of different kinds, and then Fourier transforming back \cite{Stanford2016, Patel21022017, PhysRevX.7.031047, PhysRevD.96.065005}. Because of the unique time contour, the rungs in the Bethe-Salpeter equations are correlation functions between different time folds and are called Wightman functions, given by
\begin{align}
& G_{(n,m)}^W(k, T) = \frac{A_{n,m}(k)}{2 \cosh \frac{\beta k_0}{2}}
= \frac{\pi \, \delta(k_0 + e^{m}_n(\vec k))}{\cosh \frac{\beta k_0}{2}},
\label{eq:G Wightman}
\\
\nn 
& D^W(q, T) = \frac{B(q)}{2 \sinh \frac{\beta q_0}{2}}
\\ & = \frac{q_0}{q_0^2 + \lt( c(v) (|q_x| + |q_y|) + M(T,\L,v) \rt)^2} \frac{1}{\sinh \frac{\beta q_0}{2}},
\label{eq:D Wightman}
\end{align}
where $A_{n,m}(k)$ and $B(q)$ are the fermion and boson spectral functions, respectively. The rails in the Bethe-Salpeter equation are the standard retarded Green's functions, with leading order in $w(v)$ self-energy corrections,
\begin{align}
\begin{split}
& G_{(n,m)}^R(k, T) = \[ k_0 + e^{m}_n(\vec k) 
    \rt. \\ & \lt. \hspace{5mm} + i \frac{3}{4} w(v) \, T 
	\lt(
	2 \frac{-i k_0 - \pi T }{2\pi T} 
	\log
	\left(\frac{\Lambda }{2\pi T}\right)
	\rt. \rt. \\ & \lt. \lt. \hspace{5mm} -
	2 \log \lt[
	\Gamma \left(1+\frac{- i k_0 - \pi T }{2\pi T}\right)
	\rt]
	+ \log \frac{\Lambda}{M(T,\Lambda, v)}
	\rt)
\]^{-1},
\end{split}
\label{eq:G ret}
\\
& D^R(q, T) = \[ - i q_0 + c(v) \Big[ |q_x| + |q_y| \Big] + M(T,\L,v) \]^{-1},
\label{eq:D ret}
\end{align}
and the advanced Green's functions which are simply the complex conjugate of the retarded ones, $G^{A}_{(n,m)} = G^{R \, *}_{(n,m)}, D^{A} = D^{R \, *}$. The fermion self-energy at finite $T$ is computed in Appendix \ref{sec:fermion self-energy}, and we note that to leading order it is independent of $\vec k$. To get Eq. (\ref{eq:D ret}) we simply analytically continue the propagator from Eq. (\ref{eq:2D_theory}) to real frequency, since the boson self-energy to leading order in $w$ is already given in Eq. (\ref{eq:2D_theory}). There are no other types of propagators to consider, since the interaction vertices in the expansion are only placed on the real time folds and not on the thermal circle, as it is believed those corrections will not change the spectrum of growth exponents \cite{Stanford2016}.

\section{Chaos of the fermions}

\begin{figure}[h]
	\includegraphics[width=2.5in]{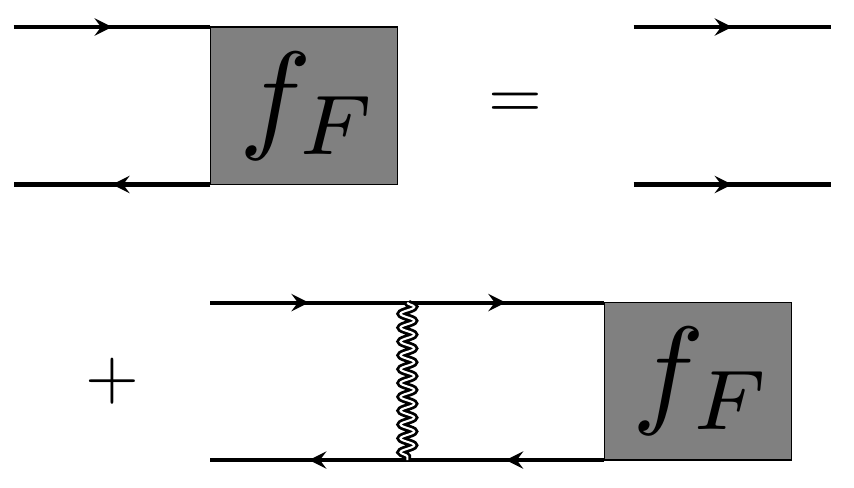}
	\caption{The leading order diagrams of the Bethe-Salpeter equation for $f_F(\omega, k)$. The first is the zero-rung diagram, while the second is the leading order rung diagram.
	}
	\label{fig:f_F Bethe-Salpeter}
\end{figure}

We start by computing the spatially averaged correlator for the fermion, $f_F(t) \equiv \int d^2 x \, f_F(t, \vec x)$. We first decompose the Fourier transform $f_F(\omega)$ as
\begin{equation}
f_F(\omega) = \f12
\sum_{n=1}^{4} \sum_{m=\pm}
\int dk
f_F^{(n,m)}(\omega, k),
\end{equation}
where $dk \equiv \frac{d^3k}{(2\pi)^3}$. We study the Bethe-Salpeter equation for each component, and then integrate over $k$ to get back $f_F(\omega)$. The Bethe-Salpeter equation for $f_F^{(n,m)}(\omega, k)$ to leading order can be expressed as
\begin{equation}
\begin{split}
f_F^{(n,m)}(\omega, k) &=  
G^R_{(n,m)}(k) \, G^{A}_{(n,m)}(k - \omega)
\\ & \hspace{-5mm} \times 
\lt[
1 +
\int dk' \,
K_F(k, k', \omega) \,
f_F^{(n,\bar m)}(\omega, k')
\rt],
\end{split}
\label{eq:f_F Bethe Salpeter 1st order}
\end{equation}
or diagrammatically in Fig. \ref{fig:f_F Bethe-Salpeter}. 
The first diagram in the series is the zero-rung diagram, which includes only self-energy corrections. The second diagram is the leading order rung contribution. 
\\
\indent 
We examine the zero-rung diagram separately. It is given by the sum over hot spots of products of a retarded and advanced Green's functions. For a given hot-spot, to leading order, we can set $v=0$ in $e_n^{m}(\vec k)$. The divergent integral over the momentum with velocity component $v$ contributes to the normalization of $f_F(t)$, but does not affect it's growth rate. We can then do the integral over the other momentum component, noting that the quantity $\log(\Lambda/M) \gg 1$ dominates the location of the pole. This gives
\begin{align}
\nn & \int dk \, f_F^{(n,m)}(\omega,k) =  
\int dk \, G^R_{(n,m)}(k) \, G^{A}_{(n,m)}(k - \omega)
\\ \nn & =
i \, \int \frac{dk_x}{2\pi} \int \frac{dk_0}{2\pi} 
\[\omega + i \frac{3}{4} w \, T 
	\lt(
	- \frac{i \omega + 2 \pi T}{\pi T}
	\log\left(\frac{\Lambda }{2\pi T}\right)
	\rt. \rt. \\ \nn & \lt. \lt. -
	2 \log \lt[
	\Gamma \left(1+\frac{- i k_0 - \pi T }{2\pi T}\right)
	\Gamma \left(1+\frac{i (k_0 - \omega) - \pi T }{2\pi T}\right)
	\rt]
	\rt. \rt. \\ & \lt. \lt. \hspace{50mm} + 2\log \frac{\Lambda}{M}
	\rt)
\]^{-1}.
\end{align}
The Fourier transform of this contribution decays exponentially in time.
\\
\indent 
The exponential growth of $f_F(t)$ must come from the rung diagram and we can therefore neglect the zero-rung diagram, which gives
\begin{equation}
\begin{split}
f_F^{(n,m)}(\omega, k) &=  
G^R_{(n,m)}(k) \, G^{A}_{(n,m)}(k - \omega)
\\ & \times \int dk' \,
K_F(k, k', \omega) \,
f_F^{(n,\bar m)}(\omega, k').
\label{eq:f_F Bethe Salpeter exp growth}
\end{split}
\end{equation}
This equation is equivalent to the statement that we can add a rung to $f_F(\omega, k)$ without changing the long-time behavior of $f_F(t)$ \cite{Stanford2016}. The kernel $K_F$ is given by
\begin{align}
\nn & K_F(k, k', \omega) =  \frac{3\pi}{2} v \, D^W (k - k')
\\ \nn & =
\frac{3 \pi}{2} v \,
\frac{k_0 - k_0'}{(k_0 - k_0')^2 + \lt( c(v) (|k_x - k_x'| + |k_y - k_y'|) + M \rt)^2}
\\ & \hspace{45mm}
\times \frac{1}{\sinh \frac{\beta (k_0 - k_0')}{2}}.
\label{eq:K 1 exact}
\end{align}
Without loss of generality, we focus on a single hot spot pair, $(1,\pm)$, for which the coupled set of equations is (for ease of notation we omit the '$1$' from the superscripts and subscripts) 
\begin{equation}
\begin{split}
f_F^{(\pm)}(\omega, k) &=  
G_{\pm}^R(k) \, G_{\pm}^{A}(k - \omega)
\\ & \times \frac{3 \pi}{2} v \int dk' \, 
D^W (k - k') \,
f_F^{(\mp)}(\omega, k').
\end{split}
\label{eq:f_F Bethe Salpeter labeled +-}
\end{equation}
As for the first diagram, we can approximate $e_1^{\pm} \approx \pm k_y$ and integrate over $k_x$. The equation for $f_F^{(\pm)}(\omega, k_0, k_y) \equiv \int \frac{d k_x}{2\pi} f_F^{(\pm)}(\omega, k)$ is given by 
\begin{widetext}
\begin{equation}
\begin{split}
f_F^{(\pm)}(\omega, k_0, k_y) &=  
\frac{1}{k_0 \pm k_y + i \frac{3}{4} w \, T 
	\lt(
	2 \frac{-i k_0 - \pi T }{2\pi T} \log
	\left(\frac{\Lambda }{2\pi T}\right)
	-
	2 \log \lt[
	\Gamma \left(1+\frac{- i k_0 - \pi T }{2\pi T}\right)
	\rt]
	+ \log \frac{\Lambda}{M}
	\rt)} 
\\ & 
\frac{1}{k_0- \omega \pm k_y - i \frac{3}{4} w \, T 
	\lt(
	2 \frac{i (k_0 - \omega) - \pi T }{2\pi T} \log
	\left(\frac{\Lambda }{2\pi T}\right)
	-
	2 \log \lt[
	\Gamma \left(1+\frac{i (k_0 - \omega) - \pi T }{2\pi T}\right)
	\rt]
	+ \log \frac{\Lambda}{M}
	\rt)}
\\
&
\times \frac{3}{4} v
\int \frac{dk_0' dk_y'}{(2\pi)^2}
\frac{\pi - 2 \arctan\lt(\frac{c(v) |k_y - k_y'| + M}{|k_0 - k_0'|}\rt)}{c(v) |k_0 - k_0'|}
\frac{k_0 - k_0'}{\sinh \frac{\beta (k_0 - k_0')}{2}}
f_F^{(\mp)}(\omega, k_0', k_y').
\end{split}
\end{equation}
Since we expect that $f_F$ depends on $k_y'$ with no small prefactor, to leading order we can set $c(v) |k_y - k_y'| \rightarrow 0$, which enables us to integrate over $k_y$ in the same way as we did for the first diagram. The equation for $f_F^{\pm}(\omega,k_0) \equiv \int \frac{dk_y}{2\pi} f_F^{\pm}(\omega,k_0,k_y)$ is entirely independent of the hot spot index and so we remove it, 
\begin{align}
\nn f_F(\omega, k_0) & =  
\[\omega + i \frac{3}{4} w \, T 
	\lt(
	- \frac{i \omega + 2 \pi T}{\pi T}
	\log\left(\frac{\Lambda }{2\pi T}\right)
	- 2 \log \lt[
	\Gamma \left(1+\frac{- i k_0 - \pi T }{2\pi T}\right)
	\Gamma \left(1+\frac{i (k_0 - \omega) - \pi T }{2\pi T}\right)
	\rt]
	+ 2\log \frac{\Lambda}{M}
	\rt)
\]^{-1}
\\ &
\hspace{60mm} \times i \frac{3}{4} v
\int \frac{dk_0'}{2\pi}
\frac{\pi - 2 \arctan\lt(\frac{M}{|k_0 - k_0'|}\rt)}{c(v) |k_0 - k_0'|}
\frac{k_0 - k_0'}{\sinh \frac{\beta (k_0 - k_0')}{2}}
f_F(\omega, k_0').
\label{eq:f_F(omega, k_0) equation}
\end{align}
We can see from the scaling that $\lambda_L^{(F)} \propto T$ up to logarithmic corrections, and we can scale out the temperature. We convert Eq. (\ref{eq:f_F(omega, k_0) equation}) to a matrix equation of the form $\mathcal{M} (\omega) f_F(\omega) = 0$,
\begin{align}
\nn &
\int \frac{dk_0'}{2\pi}
\Bigg( 
i \frac{3}{4} w(v)
\frac{\pi - 2 \arctan\lt(\frac{\bar M}{|k_0 - k_0'|}\rt)}{|k_0 - k_0'|}
\frac{k_0 - k_0'}{\sinh \frac{k_0 - k_0'}{2}} - 
2 \pi
\delta(k_0' - k_0) 
\\ & 
\times \Bigg[
\omega - i \frac{3}{2} w(v) 
\lt(
\lt(1 + \frac{i \omega}{2\pi}\rt)
\log\left(\frac{\bar \Lambda}{2\pi}\right)
+
\log \lt[
\Gamma \left(\f12 - \frac{i k_0}{2\pi}\right)
\Gamma \left(\f12 + \frac{i (k_0 - \omega)}{2\pi}\right)
\rt]
- \log\lt(\frac{\bar \Lambda}{\bar M}\rt)
\rt)
\Bigg]
\Bigg)
f_F\lt(\omega \, T, k_0' \, T \rt)
= 0,
\label{eq:matrix equation continuou f_F}
\end{align}
\end{widetext}
where $\bar \Lambda \equiv \L/T$ and $\bar M \equiv M/T$. To find the fastest exponential growth of $f_F(t)$ we look for the eigenvectors $f_F(\omega, k_0)$ of $\mathcal{M} (\omega)$ with the largest eigenvalue, for $\omega$ on the positive imaginary axis. This is done by discretizing $k_0$ in Eq. (\ref{eq:matrix equation continuou f_F}) and diagonalizing the resulting finite matrix numerically. The details of the numerical solution are outlined in Appendix \ref{sec:numerical Lyapunov exponent fermion}. We find that to the leading order, which is $\O(w)$, the Lyapunov exponent for the fermion is zero,
\begin{equation}
	\lambda_L^{(F)} = 0 + \O(w^2) \, T.
\end{equation}
Here, the $\O(w^2)$ term is understood to be up to logarithms in $w$ and $T$. Computing $\lambda_L^{(F)}$ to $\O(w^2)$ involves the computation of higher-order self-energy diagrams \cite{PhysRevX.7.021010} and the higher order rung diagrams shown in Appendix \ref{sec:higher-order graphs for f_F}, and is beyond the scope of this work. The reason for this null result is that the self energy contribution exactly cancels that coming from the kernel $K_F$. Although we don't understand the reason for this completely, it seems to be a product of the structure of the theory, the linear fermion dispersion, and the momentum-independence of the self-energy at leading order \cite{Patel21022017}. Because the Lyapunov exponent vanishes to the order we are working at, we cannot compute the spatial dependence of $f_F(t, \vec x)$ in a controlled way.  

\section{Chaos of the bosons}

\begin{figure}[h]
	\vspace{2mm}
	\includegraphics[width=3.in]{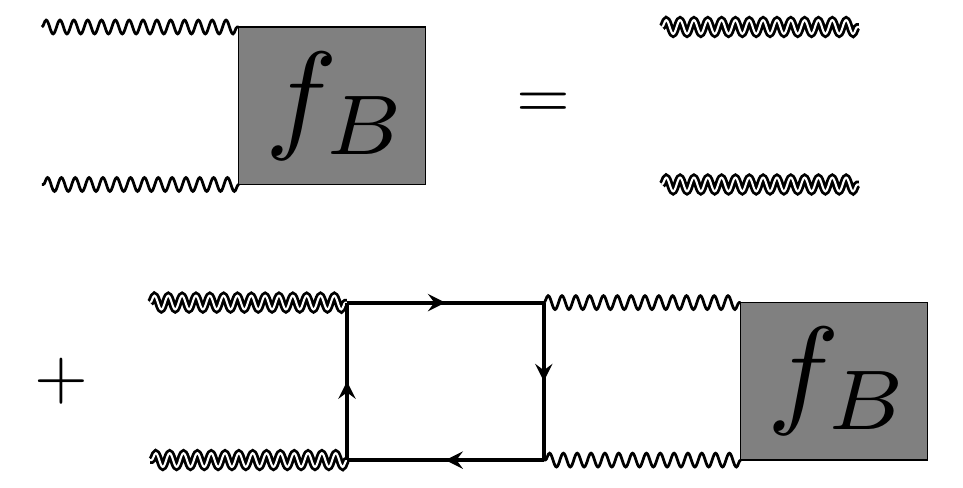}
	\caption{
		The leading order diagrams of the Bethe-Salpeter equation for $f_B(\omega, k)$. The first is the zero-rung diagram, while the second is the leading order rung diagram.
	}
	\label{fig:f_B Bethe-Salpeter simple}
\end{figure}

\begin{figure}[h]
	\includegraphics[width=3.in]{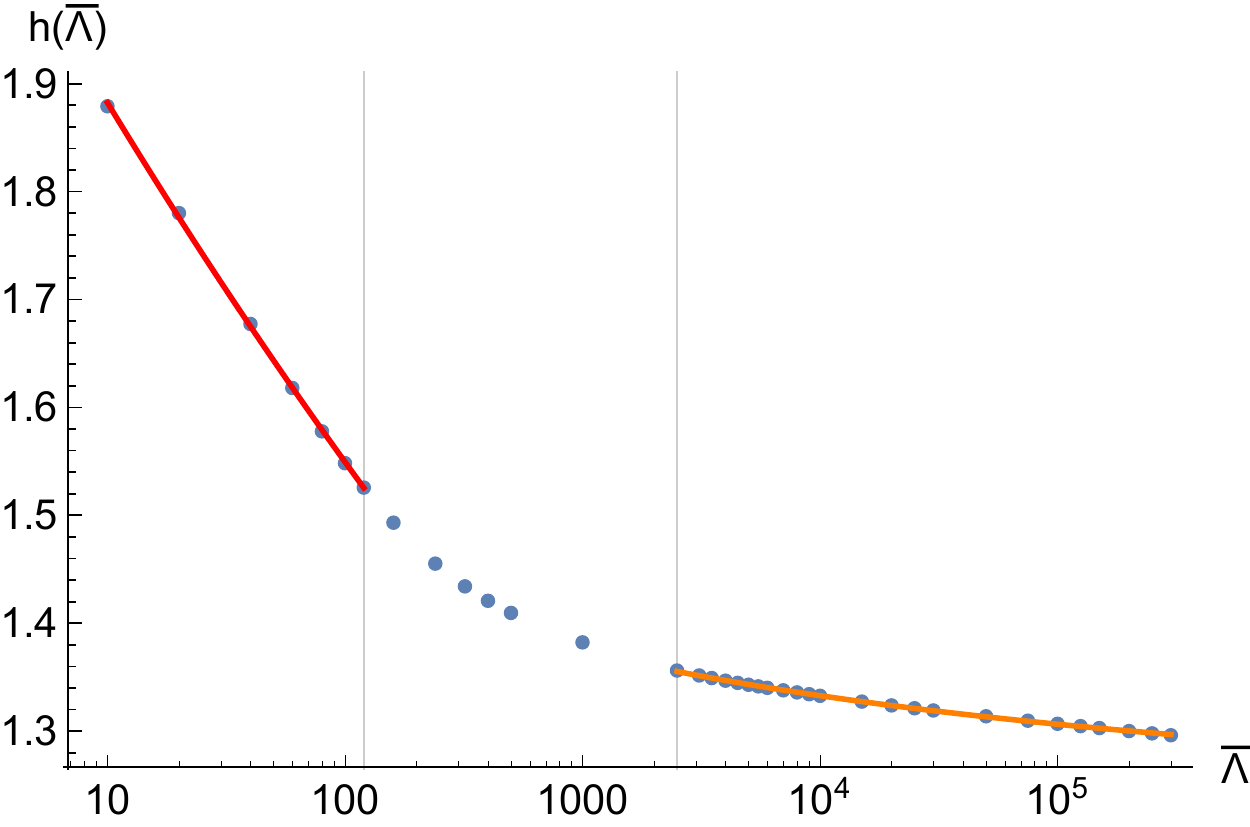}
	\caption{ The plot of $h(\bar \L)$. The two gray lines are $\bar \L = 120$ and $\bar \L = 2500$. Here, the cutoff of the integral is taken to be $15$ and the momentum spacing is $\Delta k_x = \Delta k_y= 0.25$. The fits have the same form as Eq. (\ref{eq: h fit}) with slightly different coefficients.}
	\label{fig:h of bar Lambda}
\end{figure}

For the boson, we are able to compute the full spatiotemporal dependence of $f_B(t,\vec x)$ to the leading order in $w(v)$. We start by computing $f_B(t) = \int d^2 x \, f_B(t,\vec x) = \int \frac{d \omega}{2 \pi} e^{-i \omega t} \f13 \int dk \, f_B(\omega, k)$. The Bethe-Salpeter equation for $f_B(\omega, k)$ to leading order is shown diagrammatically in Fig. \ref{fig:f_B Bethe-Salpeter simple},
and is given by
\begin{equation}
\begin{split}
f_B(\omega, k) & =  
D^R(k) \, D^{A}(k - \omega)
\\ &
\times \lt[
1 +
6 \int dk' \, 
K_B(k, k', \omega) \,
f_B(\omega, k')
\rt].
\end{split}
\label{eq:f_B Bethe Salpeter 1st order}
\end{equation}
The zero-rung contribution is trivial to compute and gives an exponential decay of $f_B(t) \propto e^{-2 \, M \, t}/(c(v)^2 t^2)$. Therefore, as for $f_F$, it can be ignored when looking for the exponential growth of $f_B(t)$. The rung contribution has no minus sign coming  from the fermion loop because two of the legs are Wightman functions. We note that there is no crossed diagram because of the time ordering properties of the expansion. The kernel $K_B$ is computed in Appendix \ref{sec:calculation of K_B} and is given by
\begin{equation}
\begin{split}
K_B(k, k', \omega)
& = \frac{i}{\omega}
\, \frac{\pi^2 v}{16} 
\, \frac{k_0 - k_0'}{\sinh \frac{\beta (k_0 - k_0')}{2}} 
\\ & \times \sum_{n=1}^{4} \sum_{m = \pm}	
\delta(k'_0 - k_0 + e_n^{(m)}(\vec{k}' - \vec k)).
\end{split}
\label{eq:K_B}
\end{equation}
Ignoring the zero-rung term we have
\begin{widetext}
\begin{equation}
\begin{split}
f_B(\omega, k) & =  
\frac{1}{ - i k_0 + c(v) \lt( |k_x| + |k_y| \rt) + M}
\,
\frac{1}{ i (k_0 - \omega) + c(v) \lt( |k_x| + |k_y| \rt) + M}
\\ & 
\times 
\frac{i}{\omega} \, \frac{6 \pi^2 v}{16}
\int dk' \, 
\frac{k_0 - k_0'}{\sinh \frac{\beta (k_0 - k_0')}{2}} 
\sum_{n=1}^{4} \sum_{m = \pm}	
\delta\lt( k'_0 - k_0 + e_n^{(m)}(\vec{k}' - \vec k) \rt) 
f_B(\omega, k').
\end{split}
\label{eq:f_B Bethe Salpeter growth}
\end{equation}
We convert Eq. (\ref{eq:f_B Bethe Salpeter growth}) to a matrix equation $\mathcal{M} (\omega) f_B(\omega) = 0$,
\begin{align}
\nn & 
\int dk_0'\, dk_x' \, dk_y'
\Bigg[
\frac{i}{\omega}  \, \frac{3 \, v}{2^5 \pi}
\frac{k_0 - k_0'}{\sinh \frac{\beta (k_0 - k_0')}{2}} 
\sum_{\pm}
\Bigg(
\delta(k'_0 - k_0 \pm (k_y' - k_y))
+
\delta(k'_0 - k_0 \pm (k_x' - k_x))
\Bigg) 
\\ & -
\delta(k' - k)
\[- i k_0 + c(v) \( |k_x| + |k_y| \) + M \] \[ i (k_0 - \omega) + c(v) \( |k_x| + |k_y| \) + M \] 
\Bigg]
f_B(\omega, k') = 0,
\label{eq:f_B Bethe Salpeter calculation 2}
\end{align}
where we have set $v=0$ in the fermion dispersions, since it is a subleading contribution. As for the fermion, the fastest exponential growth of $f_B(t)$ will be given by the largest eigenvalue of $\mathcal{M}(\omega)$ for $\omega$ on the positive imaginary axis. We have to find this eigenvalue numerically, by discretizing the integration variables. We scale out temperature and work with dimensionless units of frequency and momentum. The matrix $\mathcal{M}_{3\text{D}}(\omega)$ to diagonalize is given by
\begin{align}
\nn & 
\mathcal{M}_{3\text{D}}(\omega) = 
\Bigg[
\Delta k'_x \Delta k'_y \, 
\frac{i}{\omega}
\frac{3 \, v}{2^5 \pi}
\frac{k_0 - k_0'}{\sinh \frac{k_0 - k_0'}{2}} 
\Bigg(
\delta_{k'_0, k_0 + k_y' - k_y}
+
\delta_{k'_0, k_0 - k_y' + k_y}
+ 
\delta_{k'_0, k_0 + k_x' - k_x}
+
\delta_{k'_0, k_0 - k_x' + k_x}
\Bigg) 
\\ & \qquad \qquad  -
\delta_{k'_0, k_0} \, \delta_{k'_x, k_x} \, \delta_{k'_y, k_y}
\[- i k_0 + c(v) \( |k_x| + |k_y| \) + \bar{M} \] \[ i (k_0 - \omega) + c(v) \( |k_x| + |k_y| \) + \bar{M} \] 
\Bigg].
\label{eq:M v=0 discretized}
\end{align}
By solving this equation for small system sizes we can see that the eigenvector $f_B(i \lambda_L^{(B)})$ of the largest eigenvalue is almost a delta function peak at $k_0 = 0$. We can therefore use the ansatz $f_B(\omega, k) = g_B(\omega, \vec k) \delta(k_0)$. Then, we can integrate both sides of the equation over $k_0$, which uses the delta functions to fix $k_0-k_0'$. This gives an integral equation for $g_B(\omega, \vec k)$, for which the new matrix is
\begin{equation}
\mathcal{M}_{2\text{D}}(\omega) = 
\Bigg[
\Delta k_x \, \Delta k_y
\frac{i}{\omega} 
\frac{3 \, v}{2^4 \, \pi}
\Bigg(
\frac{k_y - k_y'}{\sinh \frac{k_y - k_y'}{2}} 
+
\frac{k_x - k_x'}{\sinh \frac{k_x - k_x'}{2}} 
\Bigg)
-
\delta_{k'_x, k_x} \, \delta_{k'_y, k_y}
\[c(v) \( |k_x| + |k_y| \) + \bar{M} \] \[ - i \omega + c(v) \( |k_x| + |k_y| \) + \bar{M} \] 
\Bigg].
\label{eq:M 2D}
\end{equation}
\end{widetext}
$\mathcal{M}_{2\text{D}}(\omega)$ is in one lower dimension, and is therefore computationally more manageable. The details of the numerical computation are in Appendix \ref{sec:calculation of lambda_L^B}. The scaling dictates that $\lambda_L^{(B)} \propto T$ up to logarithmic corrections. From analyzing the numerically obtained solution $\lambda_L^{(B)}(v,\bar \L)/T$ ($\bar \L \equiv \Lambda/T$), we can see that for a fixed $\bar \L$ the dependence on $v$ is well fit by
\begin{equation}
	\lambda_L^{(B)}(v,\Lambda,T) = T \, h(\bar \L) \, \sqrt{w(v)} \( \log \frac{1}{w(v)} \)^{1/4},
	\label{eq:Lyapunov exponent boson}
\end{equation}
with $h(\bar \L)$ plotted in Fig. \ref{fig:h of bar Lambda}.  
We can fit the small and large $\bar \L$ regions to an inverse logarithmic form, $\frac{\alpha}{\log(\bar \L) + \beta} - \gamma$. From extrapolating to the infinite cutoff (here the cutoff of the integral is taken to be different from $\bar \Lambda$, see Appendix \ref{sec:calculation of lambda_L^B} for details) and thermodynamic limits we find the coefficients to be
\begin{equation}
h(\L/T) \approx 
\begin{cases}
\frac{150.1}{\log(\bar \L) + 30.8} - 3.1 & \bar \L \lesssim 120 \\
\text{crossover} & 120 \lesssim \bar \L  \lesssim 2500 \\
\frac{2.28}{\log(\bar \L) + 3.36} + 1.39 & \bar \L  \gtrsim 2500.	
\end{cases}
\label{eq: h fit}
\end{equation}
\\
\indent 
The result of Eq.(\ref{eq:Lyapunov exponent boson}) is larger than the expected $\O(w)$ by a factor of $w^{-1/2}$. This enhancement comes from the thermal mass, which plays an important role: without it $\lambda_L^{(B)}$ would be infinite. This is qualitatively similar to Ref. \cite{Stanford2016}, where at criticality the thermal mass changed both the scaling with coupling and with temperature of the Lyapunov exponent. However, here the scaling of $\lambda_L^{(B)}$ with small $M$ is not as straightforward. Despite this enhancement, $\lambda_L^{(B)}$ still vanishes in the small $v$ limit, which indicates that as the effective velocity of the boson $c(v) \rightarrow 0$, the $\O(1)$ Landau damping is not enough to induce chaos.

\subsection{Propagation of chaos in space}

\begin{figure}[!ht]
	\subfigure[]
	{\includegraphics[width=3.in]{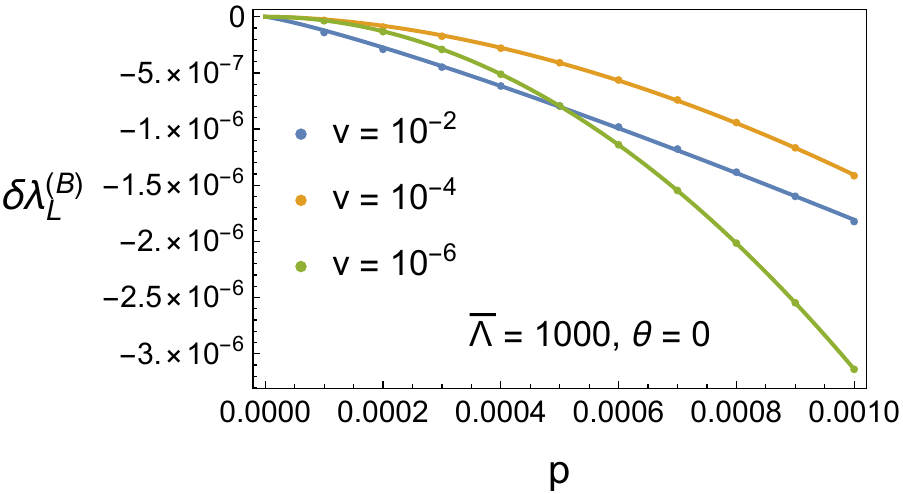}
	    } ~~~~~
	\\
	\subfigure[]
	{\includegraphics[width=2.5in]{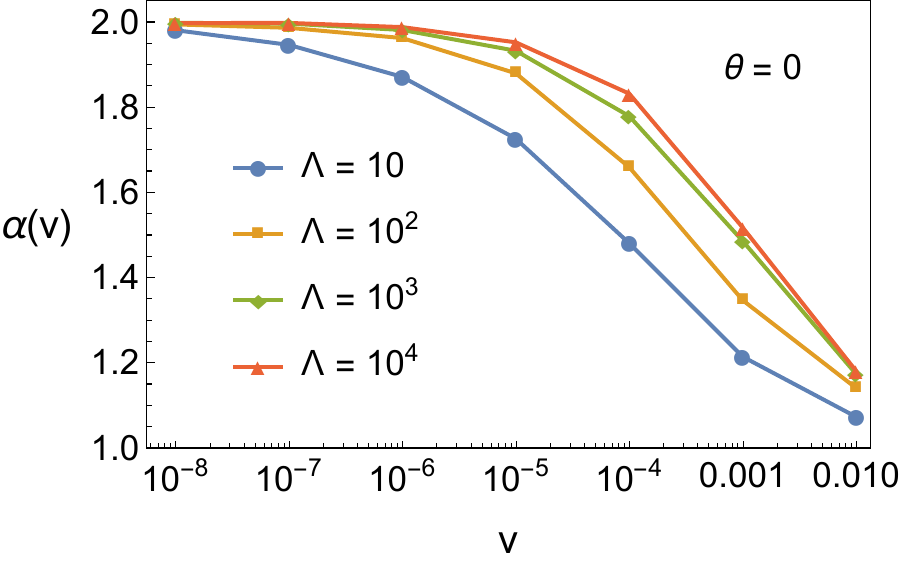}
	    } ~~~~~
	\subfigure[]
	{\includegraphics[width=2.5in]{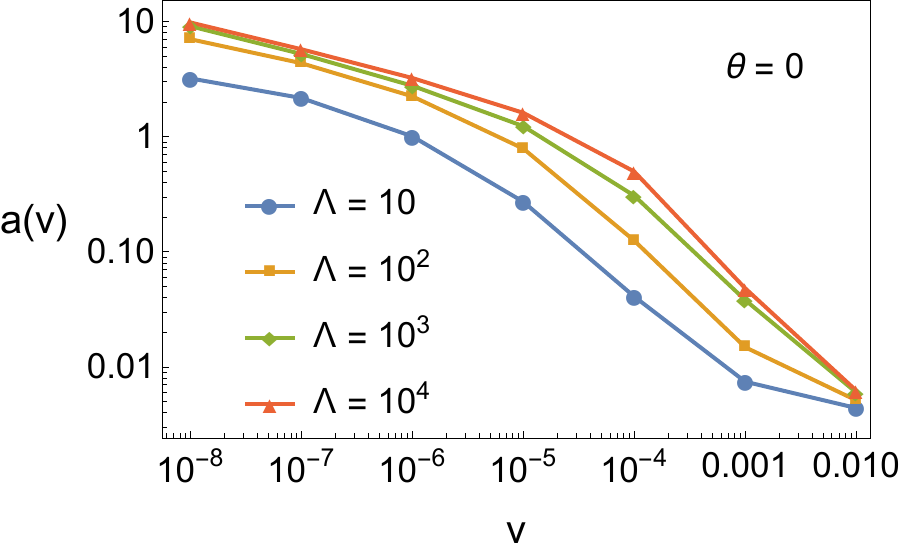}
	    } ~~~~~
	\caption{(a): Plots of the deviation $\delta \lambda_L^{(B)}(p,\theta) = - T \, a(v, \bar \L, \theta) \, p^{\alpha(v, \bar \L, \theta)}$ for various decreasing $v$ and fixed $\bar \L = 1000$ and $\theta = 0$. (b),(c): The form of $\alpha(v,\bar \L,\theta=0)$ and $\a(v,\bar \L,\theta=0)$ as functions of $v$ for various increasing $\bar \Lambda$. Between $10 < \bar \L < 10^4$ $\alpha$ increases by a factor of less than $1.5$ and $a$ increases by a factor of at most $\sim 5$.
	Since we do not fit the functional forms in these plots, there is no need in a large cutoff and small momentum-spacing scaling analysis.}
	\label{fig:finite p plots}
\end{figure}

\begin{figure}[h]
	\includegraphics[width=2.in]{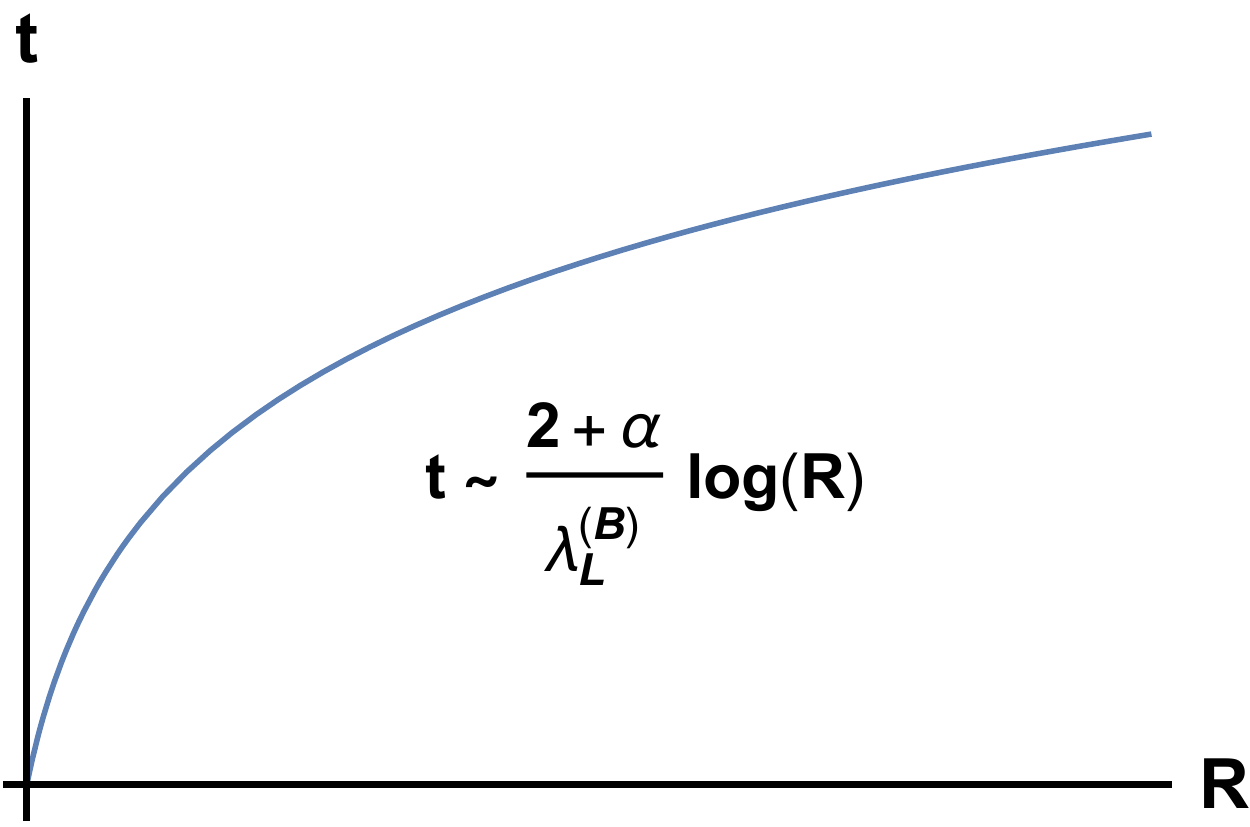}
	\caption{The exponential growth of the region where $f_B(t,\abs{R(t)}) \sim 1$ has been reached. Since there is no linear light cone, the butterfly velocity is formally infinite.}
	\label{fig:R of t}
\end{figure}

To compute the spatial dependence of $f_B(t, \vec x)$ we need to solve the same Bethe-Salpeter equation as in Fig. \ref{fig:f_B Bethe-Salpeter simple}, but with an external momentum $\vec p$ injected into the correlation function, $f_B(\omega, \vec p, k)$. By the same reasoning as for $\vec p = 0$ (see Appendix \ref{sec:calculation of lambda_L^B}) we arrive at the resulting two-dimensional matrix,
\begin{widetext}
\begin{equation}
\begin{split}
\mathcal{M}_{2\text{D}}(\omega, \vec p) = 
\Bigg[
&
\Delta k_x \, \Delta k_y
i
\frac{3 \, v}{2^4 \, \pi}
\Bigg(
\frac{k_y - k_y'}{\sinh \frac{k_y - k_y'}{2}} 
\frac{\omega}{\omega^2 - p_y^2}
+
\frac{k_x - k_x'}{\sinh \frac{k_x - k_x'}{2}} 
\frac{\omega}{\omega^2 - p_x^2}
\Bigg)
\\ & -
\delta_{k'_x, k_x} \, \delta_{k'_y, k_y}
\[c(v) \( |k_x| + |k_y| \) + \bar{M} \] \[ - i \omega + c(v) \( |k_x - p_x| + |k_y - p_y| \) + \bar{M} \] 
\Bigg].
\end{split}
\label{eq:M 2D external p}
\end{equation}
\end{widetext}
Finding the largest eigenvalue of $\mathcal{M}_{2\text{D}}(\omega, \vec p)$ for small $\vec p$ gives us the deviation
\begin{equation}
\delta \lambda_L^{(B)}(p,\theta) = - T \, a(v, \bar \L, \theta) \, p^{\alpha(v, \bar \L, \theta)},
\label{eq:Lyapunov exponent small p}
\end{equation}
where $\vec p = (p \cos(\theta), p \sin(\theta))$. We note that $\lambda_L^{(B)}(p,\theta)$ doesn't develop an imaginary part. The exponent $\alpha$ varies between $1 < \alpha < 2$, starting out close to one for larger $v$ and monotonically approaching two as $v$ approaches zero. The coefficient $a$ grows monotonically as $v \rightarrow 0$, eventually scaling as $a(v) \sim v^{-1/4}$ when $v$ is small enough (``small enough'' depends on the value of $\bar \L$). The dependence of $\alpha$,$a$ on $\theta$ is  weak. The dependence of $\alpha$,$a$ on $\bar \L$ is logarithmic, and in the range we study $10 < \bar \L < 10^5$ there is not much of a change. The plots of $a(v, \bar \L, \theta)$ and $\alpha(v, \bar \L, \theta)$ as functions of $v$ for various values of $\bar \L$ are shown in Fig. \ref{fig:finite p plots} for $\theta = 0$. More plots detailing the $\theta$ dependence of the parameters are shown in Appendix \ref{sec:calculation of lambda_L^B}.
\\
\indent
The full form of the leading time dependence of $f_B(t,\vec p)$ is $f_B(t,\vec p) \sim e^{(\lambda_L^{(B)} - a \, p^{\alpha}) \, t} \tilde f_B(t, \vec p)$, where $\tilde f_B(t, \vec p)$ is the eigenvector of the dominant eigenvalue, and we have again set $T=1$. The $\vec{p}$ dependence of the eigenvector in certain situations is known to have singularities for complex $\vec{p}$ that modify the real space structure of scrambling at large $|\vec{x}|$ and $t$ \cite{Gucci,GuKit,GuoGu}. Namely, there could exist a complex $\vec{p}$ for which $\lambda_L^{(B)}(\vec{p})$ is at its maximal value of $2\pi$ and $\tilde{f}_B$ is singular. However, Ref. \cite{GuoGu} found that (i) such a singularity in $\tilde{f}_B$ does not occur for the OTO correlation functions of the form in Eq. (\ref{eq:f_B definition}) that we compute (``retarded" OTO correlators in the language of Ref. \cite{GuoGu}), and, independently, (ii) if $\lambda_L^{(B)}(\vec{p}=0)\ll 2\pi$, as is the case in this paper, the singularity in the amplitude $\tilde{f}_B$, if it occurs, would occur at large values of complex $\vec{p}$ that would render the contribution to the Fourier transform of $f_B$ from those $\vec{p}$ severely suppressed by the exponential factor for most values of $\vec{x},t$. Therefore, it is safe for us to disregard the $\vec{p}$ dependence of $\tilde{f}_B$. 
\\ \indent 
Computing the Fourier transform of the exponential factor then gives
\begin{equation}
f_B(t, \vec x) = e^{\lambda_L^{(B)} t} \int d\theta \, dp \, p \, e^{i \, p \, \abs{\vec x}  \cos(\theta - \theta_x)} 
e^{- a(v, \bar \L, \theta) \, p^{\a(v, \bar \L, \theta)} \, t},  
\end{equation}
where $\theta_x$ is the angle of $\vec x$. Since the dependence of $a$ and $\a$ on $\theta$ is weak, we can ignore it, as the qualitative $\abs{\vec x}$ dependence will not be affected. Integrating over $\theta$ in this approximation gives, 
\begin{equation}
\begin{split}
f_B(t, \abs{\vec x}) & \sim 
e^{\lambda_L^{(B)} t} \frac{1}{(a(v, \bar \L) \, t)^{2/\alpha(v, \bar \L)}} 
\\ & \times \int dp \, p \, J_0\(p \frac{\abs{\vec x}}{(a(v, \bar \L) \, t)^{1/\alpha(v, \bar \L)}}\) 
\, e^{-p^{\a(v, \bar \L)}},
\end{split}
\end{equation}
where $J_0$ is the zeroth Bessel function of the first kind, and we have scaled out $a(v, \bar \L) \, t$ from the $p$ integral. This last integral is hard to compute analytically, but from numerics we can see that for $\abs{\vec x} \gtrsim 10 \, (a \, t)^{1/\alpha}$ the integration gives an inverse power-law scaling of  
\begin{equation}
f_B(t, \abs{\vec x})
\sim e^{\lambda_L^{(B)} t} 
\frac{a(v, \bar \L) \, t}{\abs{\vec x}^{2+\a(v, \bar \L)}}.
\label{eq:f_B with numerical x scaling}
\end{equation}
The proportionality constant is fairly flat until $\alpha$ is extremely close to $2$, and since we are studying small but finite $v$ we treat the proportionality constant as effectively $\alpha$-independent. 
\\
\indent 
From Eq. (\ref{eq:f_B with numerical x scaling}) we can define a typical ``operator radius'' for the boson operator, which is defined as the $R(t)$ for which $f_B(t, R(t)) \sim 1$ (here we forget about the angular dependence as it is weak). This is the radius within which the initially local operator at $\vec x = 0$ has spread, and is given by
\begin{equation}
R(t) \sim 
e^{\lambda_L^{(B)} t/(2 + \a(v, \bar \L))} 
(a(v, \bar \L) \, t)^{\frac{1}{2 + \a(v, \bar \L)}}.
\label{eq:R(t)}
\end{equation}
The form of Eq. (\ref{eq:R(t)}) tells us how fast the scrambling of the boson operators spreads through the system. For large $R(t)$, which we are interested in, the ``scrambled'' region grows exponentially with time, as illustrated in Fig. \ref{fig:R of t}. 
Since the boundary of this region cannot be linearized, we cannot define a finite butterfly velocity. The reason for this strange behavior is due to the highly non-local nature of the boson propagator: $D^{R,-1}(t,\vec{x}) = \pi^{2} e^{M \, t} (1 + (\frac{x_1}{c(v) \, t})^2) (1 + (\frac{x_2}{c(v) \, t})^2)$. The algebraic decay of $D^R(t,\vec x)$ with $\abs{\vec x}$ for any non-zero $t$ means that the boson is highly delocalized, and therefore an initial perturbation at the origin propagates with ``infinite'' velocity. As $c(v) \sim \sqrt{v} \rightarrow 0$, the non-locality in the propagator (slowly) disappears and instead the boson becomes completely localized in space. In the limit $v \rightarrow 0$, $\alpha(v) \rightarrow 2$ and Eq. (\ref{eq:R(t)}) is modified to $R(t) \sim 2 \sqrt{a \, \lambda_L^{(B)}} \, t$. In this case, after restoring factors of $v_F$, the butterfly velocity  would be $v_B^{(B)} = 2 \sqrt{a \, \lambda_L^{(B)}} \, v_F$, which is independent of $v$ since the $v$-dependence from $a$ and $\lambda_L^{(B)}$ would cancel. This again makes sense, since the only way a completely localized boson can propagate is by fermionic particle-hole production, which travel at a speed of $v_F$. We note, though, that the complete localization of the boson for $c(v) \rightarrow 0$ is not real, since at extremely small $c(v)$ the bare term $c_0^2 \abs{\vec q}^2/\tilde \L$ would control the boson dynamics.

\section{Discussion}

We have computed the Lyapunov exponents that describe the growth rate of both fermion and boson OTO correlators. We worked at the leading perturbative order in the control parameter $w$, and naively it seemed that both exponents might be $\sim \O(w)$. Instead, it turns out that the boson scrambles significantly faster than the fermion, $\lambda_L^{(B)} \sim \O(\sqrt{w}) \, T \gg \lambda_L^{(F)} \sim \O(w^2) \, T$. This large discrepancy stems from their different degrees of renormalization. The reason for $\lambda_L^{(F)}$ being smaller than $\O(w)$ seems to be accidental, and not a generic feature of Yukawa-type field theories. The enhancement of $\lambda_L^{(B)}$, however, is not accidental and is due to the irrelevance of the bare terms in the bosonic propagator. However, even though the Landau-damped frequency dependence of the boson kinetic term is $\O(1)$, the degree of chaos still vanishes in the $v \rightarrow 0$ limit. This indicates that strong Landau damping alone is not enough to rapidly scramble the system. Even though this fixed point has an effective coupling of order unity that controls the low-energy boson dynamics, the degree of non-integrability of the boson is still proportional to a positive power of $w$, and its scrambling rate is not close to the maximal one.
\\ \indent 
We also find that the boson scrambling spreads in space via a logarithmic light cone and an infinite butterfly velocity. Initially, one would expect that this is at odds with the Lieb-Robinson bound \cite{LiebRobinson} on the spread of information in quantum systems. However, these features have been seen before in systems with long-range interactions \cite{Hastings2006, PhysRevLett.113.030602, PhysRevLett.114.157201}, which violate the assumptions set forth in the Lieb-Robinson bound. Here, the effective action of Eq. (\ref{eq:2D_theory}) violates those same assumptions because of the non-local boson propagator, and our results are consistent with recent findings.

\section*{Acknowledgments}

We thank Igor Aleiner, Antoine Georges, Martin Claassen, Vadim Oganesyan and Subir Sachdev for useful discussions. The Flatiron Institute is a division of the Simons Foundation. AAP was supported by the US Department of Energy under Grant No. DE-SC0019030, and by a Harvard-GSAS Merit Fellowship.


\bibliographystyle{apsrev4-1}
\bibliography{sdw}

\appendix
\widetext

\setcounter{page}{1}
\setcounter{figure}{0}
\setcounter{subfigure}{0}
\setcounter{table}{0}
\setcounter{equation}{0}


\renewcommand\thefigure{A\arabic{figure}}
\renewcommand\thetable{A\arabic{table}}
\renewcommand\theequation{A\arabic{equation}}

\section{Thermal mass}
\label{sec:thermal mass}

Here we compute the thermal mass $M(T,\L,v)$ to the leading non-vanishing order in $w(v)$. At the leading order of $\O(w \log(w))$ it was computed in Ref. \cite{PhysRevX.7.021010} and found to vanish. However, we expect is to have a non-zero value at $O(w)$. The only contribution at this order comes from the diagram in Fig. \ref{fig:2LSEb}. Is it given by
\begin{equation}
M = \Pi^{(2L)}(0,0,T) - \Pi^{(2L)}(0,0,0),
\end{equation}
where 
\begin{equation}
\begin{split}
\Pi^{(2L)}(0,0,T) & = - \frac{\pi^2 v^2}{2} T \sum_{\omega_p} \int \frac{d\vec{p}}{(2\pi)^2}  D(\omega_p, \vec p, T)
\\ & 
T \sum_{\omega_k} \int \frac{d\vec{k}}{(2\pi)^2} \sum_{n=1}^{4} \sum_{m=\pm} G_{n,m}(\omega_k,\vec k) 
G_{n,\bar m}(\omega_k +\omega_p,\vec k + \vec p) 
G_{n,m}(\omega_k +\omega_p,\vec k + \vec p)
G_{n,\bar m}(\omega_k,\vec k),
\end{split}
\end{equation}
with $\omega_p, \omega_k$ being bosonic and fermionic Matsubara frequencies, respectively, and the zero temperature polarization is the straightforward analogue. We first change variables to $k_{+} = e_n^{(m)}(\vec k), k_{-} = e_n^{(\bar m)}(\vec k)$ (which has a Jacobian of $1/2v$). The integrations over $k_{\pm}$ are done via poles. The Matsubara sum over $\omega_k$ is straightforward and is equal to the integral at zero temperature. This bring us to
\begin{equation}
M(T,\L,v) = \frac{\pi v}{8} \sum_{n=1}^{4} \sum_{m=\pm} 
\int \frac{d\vec{p}}{(2\pi)^2}
\(
T \sum_{\omega_p}
\frac{\abs{\omega_p}}{\abs{\omega_p} + \eps(\vec p, M)} 
-
\int \frac{d\omega_p}{2\pi}
\frac{\abs{\omega_p}}{\abs{\omega_p} + \eps(\vec p)} 
\)
\frac{1}{(i \omega_p + e_n^{(m)}(\vec p))(i \omega_p + e_n^{(\bar m)}(\vec p))},
\end{equation}
where $\eps(\vec p,M) = c(\abs{p_x} + \abs{p_y}) + M$ and $\eps(\vec p) = c(\abs{p_x} + \abs{p_y})$. The Matsubara summation and the frequency integration can be done directly, which gives
\begin{equation}
\begin{split}
M(T,\Lambda,v) &= \frac{\pi v}{8 (2 \pi)^3} \int_{-\Lambda}^{\Lambda} dp_x dp_y \sum_{n=1}^{4} \sum_{m=\pm} 
\Bigg[
-\frac{2 \eps(\vec p, M) \(\eps(\vec p, M)^2- e_n^{(m)}(\vec p) \, e_n^{(\bar m)}(\vec p) \) \psi \(\frac{\eps(\vec p, M)}{2 \pi T}\)}{\(\eps(\vec p, M)^2+e_n^{(m)}(\vec p)^2\)
	\(\eps(\vec p, M)^2 + e_n^{(\bar m)}(\vec p)^2\)}
\\ & +
\frac{1}{e_n^{(m)}(\vec p) - e_n^{(\bar m)}(\vec p) }
\(
\frac{e_n^{(m)}(\vec p) \, \psi \(1+i \frac{e_n^{(m)}(\vec p)}{2 \pi T}\)+2 \pi i T}
{\eps(\vec p, \bar M)-i e_n^{(m)}(\vec p)}
+
\frac{e_n^{(m)}(\vec p) \, \psi \(-i \frac{e_n^{(m)}(\vec p)}{2 \pi T}\)}
{\eps(\vec p, M)+i e_n^{(m)}(\vec p)}
\rt.
\\ & -
\lt.
\frac{e_n^{(\bar m)}(\vec p) \,  \psi \(1 + i \frac{e_n^{(\bar m)}(\vec p)}{2 \pi T}\)+2 \pi i T}
{\eps(\vec p, M)-i e_n^{(\bar m)}(\vec p)}
-
\frac{e_n^{(\bar m)}(\vec p) \, \psi \(-i \frac{e_n^{(\bar m)}(\vec p)}{2 \pi T}\)}
{\eps(\vec p, M)+i e_n^{(\bar m)}(\vec p)}
\rt.
\\ & +
\lt.
\frac{e_n^{(m)}(\vec p) \( \pi \, \abs{e_n^{(m)}(\vec p)} + 2 \, \eps(\vec p) \log\(\frac{\eps(\vec p)}{\abs{e_n^{(m)}(\vec p)}} \) \) }
{\eps(\vec p)^2 + (e_n^{(m)}(\vec p))^2}
-
\frac{e_n^{(\bar m)}(\vec p) \( \pi \, \abs{e_n^{(\bar m)}(\vec p)} + 2 \, \eps(\vec p) \log\(\frac{\eps(\vec p)}{\abs{e_n^{(\bar m)}(\vec p)}} \) \)}
{\eps(\vec p)^2 + (e_n^{(\bar m)}(\vec p))^2}
\)
\Bigg],
\end{split}
\label{eq:numerical M full}
\end{equation}
where $\psi(z) = \Gamma'(z)/\Gamma(z)$ is the digamma function. The thermal mass is proportional to temperature (up to logarithms), as can be seen from scaling $T$ out of Eq. (\ref{eq:numerical M full}). Since the only two scales are $T$ and $\Lambda$, the ratio $M(T,\L,v)/T$ is a function of $v$ and $\bar \Lambda  = \L/T$ only. We can therefore set $T = 1$ and compute $\bar M(v,\bar \L) \equiv M(v,\bar \Lambda)/T$.
\\
\indent 
We solve Eq. (\ref{eq:numerical M full}) numerically for $\bar \Lambda \geq 10$ and $v \leq 10^{-2}$. For a fixed $\bar \Lambda$ the scaling with $v$ is hard to fit systematically. We therefore fix $v$ and fit the resulting curves as functions of $\bar \L$. These curves are shown in Fig. \ref{fig: M(v,Lambda)}. 
\begin{figure}
	\centering
	\subfigure[\, $v = 10^{-2}$]
	{\includegraphics[scale=0.5]{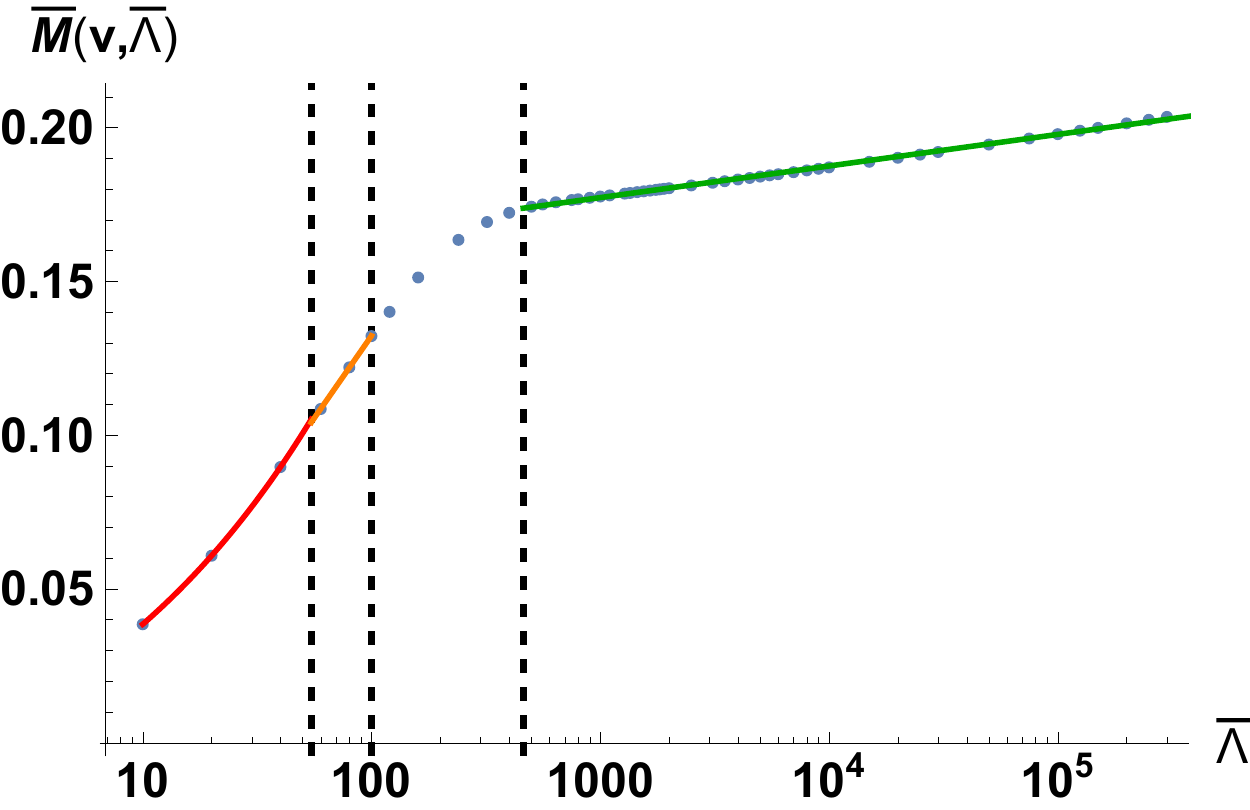}
		} 
	\subfigure[\, $v = 10^{-3}$]
	{\includegraphics[scale=0.5]{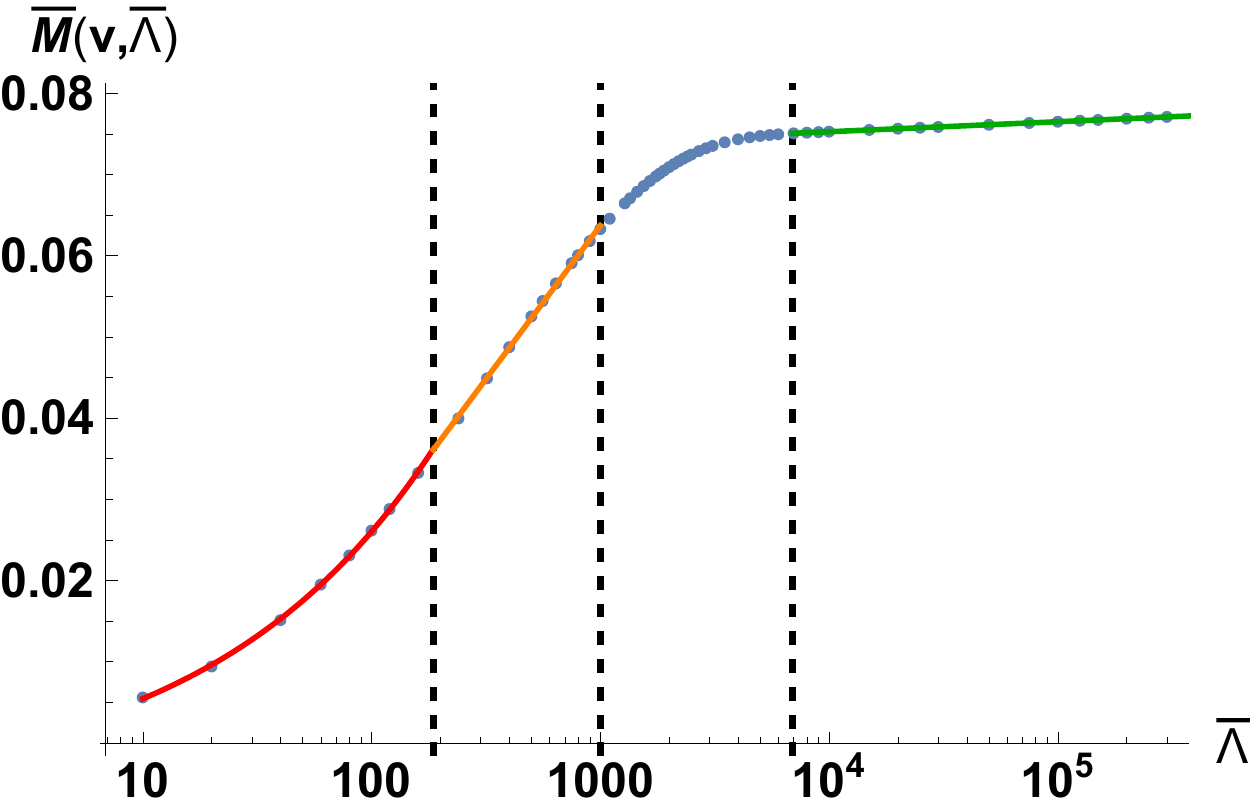}
		} 
	\subfigure[\, $v = 10^{-4}$]
	{\includegraphics[scale=0.5]{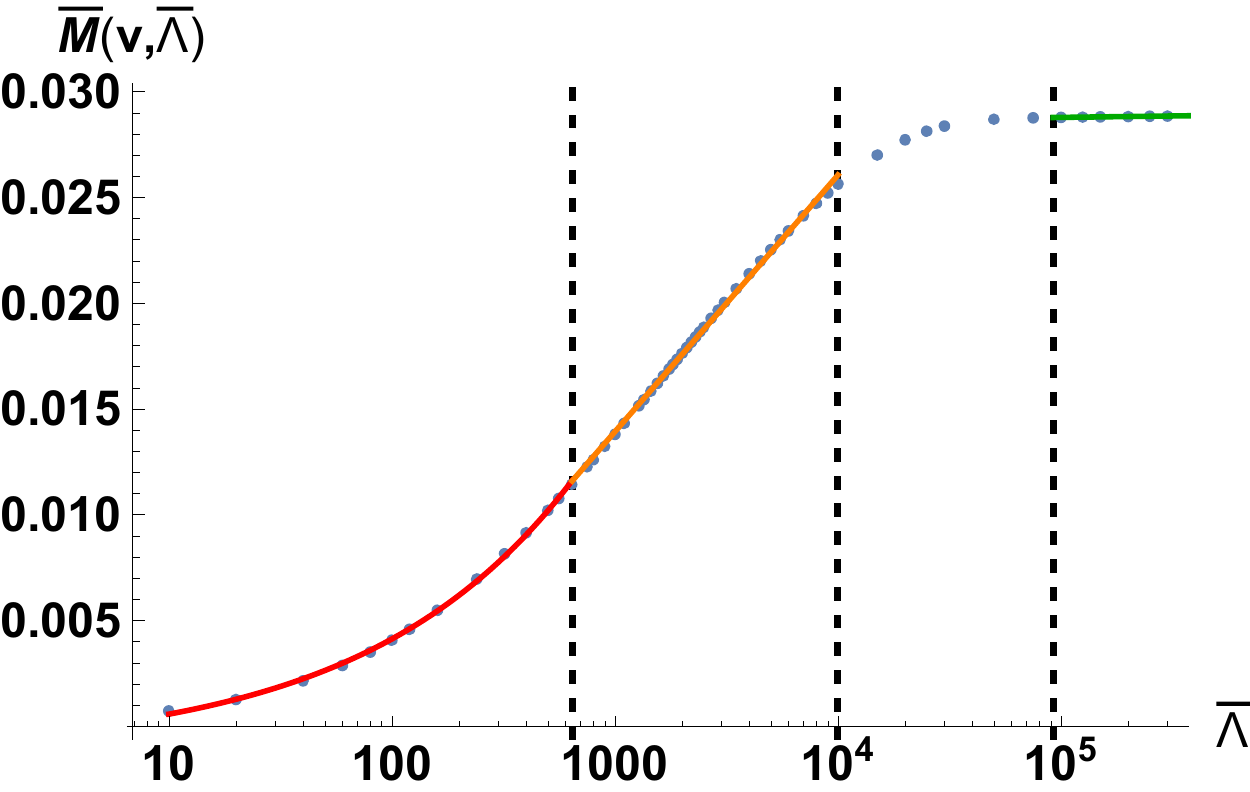}
		} 
	\subfigure[\, $v = 10^{-5}$]
	{\includegraphics[scale=0.5]{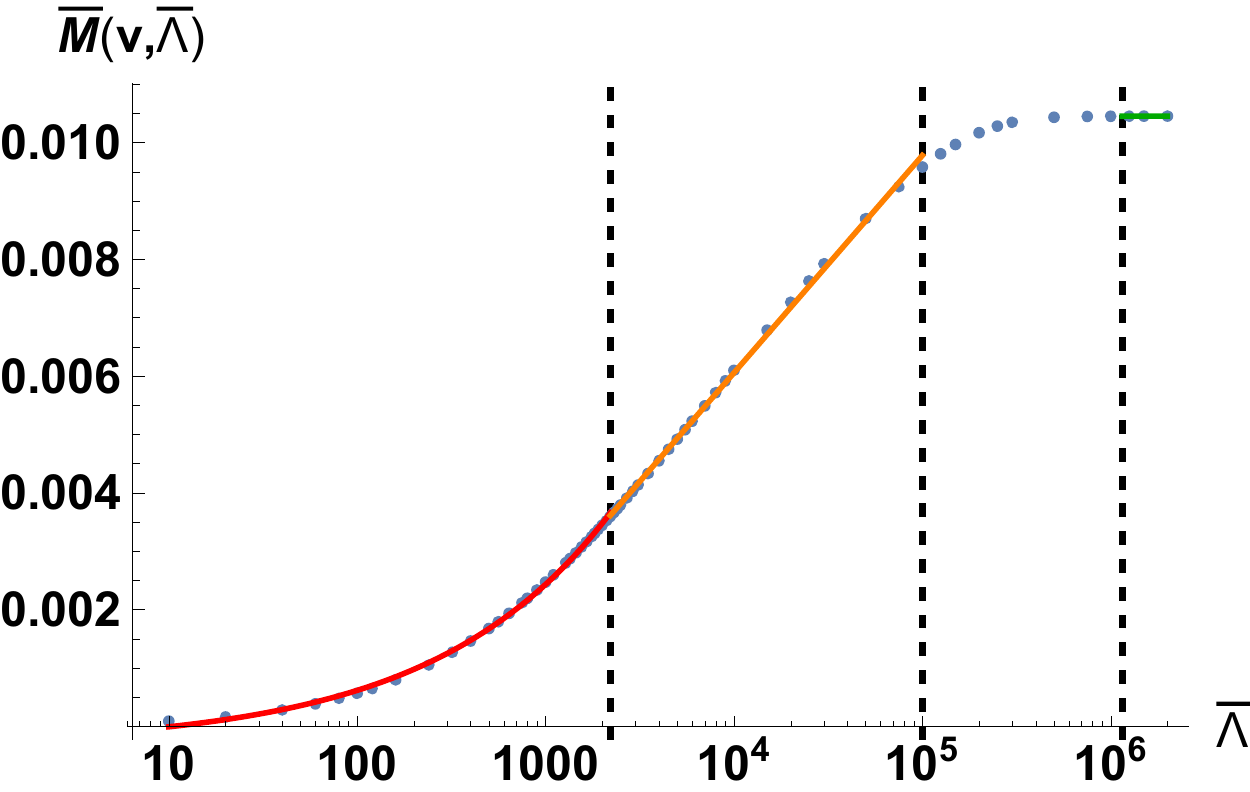}
		} 
	\caption{
		The solution of Eq. (\ref{eq:numerical M full}) as a function of $\bar \L$ for various values of $v$, plotted on a log-linear scale. The three dashed lines in each plot mark the points where $\bar \Lambda = \frac{1}{c} \log\frac{1}{c}, \frac{1}{v}$, and $\frac{1}{v} \log\frac{1}{v}$, in increasing order.
	}
	\label{fig: M(v,Lambda)}
\end{figure}
From fitting the various regions and values at their junctions we find that 
\begin{equation}
\bar M(v,\bar \Lambda) = 
\begin{cases}
a_0(v) \sqrt{\bar \Lambda} + b_0(v) & 10 < \bar \Lambda < \frac{1}{c} \log\frac{1}{c}\\
3.4 \frac{c}{\sqrt{\log \frac{1}{c} }} & \bar \L = \frac{1}{c} \log \frac{1}{c} \\
a_1(v) \log \bar \Lambda + b_1(v) & \frac{1}{c} \log\frac{1}{c} < \bar \Lambda < \frac{1}{v}\\
1.5 \, c \sqrt{\log \frac{1}{c} } & \bar \L = \frac{1}{v} \\
\text{crossover} & \frac{1}{v}  < \bar \Lambda < \frac{1}{v} \log\frac{1}{v}\\
 3.7 \, c & \bar \L = \frac{1}{v}  \log \frac{1}{v}\\
a_2(v) \log \bar \Lambda + b_2(v) & \frac{1}{v} \log\frac{1}{v} < \bar \Lambda.
\end{cases}
\label{eq:numerical M full solution}
\end{equation}
The functions $a_i(v)$ and $b_i(v)$ are well estimated by 
\begin{align*}
a_0(v) & \approx v^{3/4} \, \( 0.55 - \frac{0.18}{\log\frac{1}{v}} \),~~b_0(v)  \approx v^{3/4} \, \( 0.11 \, \log\frac{1}{v} - 0.13 \), \\
a_1(v) & \approx v^{1/2} \( 0.62 - \frac{0.72}{\log\frac{1}{v}} \), 
\quad b_1(v) \approx - 0.22 \, v^{1/2} \, \log\frac{1}{v}, \\
a_2(v) & \approx v \( 0.72 - \frac{1.25}{\log\frac{1}{v}} \), 
\qquad b_2(v) \approx 0.32 \, v^{1/2} \, \log\frac{1}{v}.
\label{eq:a_i estimates}
\end{align*}
For a fixed $\bar \Lambda$, as we take $v$ to be smaller, $\bar M(v, \bar \L)$ changes between the three different regimes indicated in \eq{eq:numerical M full solution}, and therefore the scaling with $v$ is not uniform for all $v$. 


\renewcommand\thefigure{B\arabic{figure}}
\renewcommand\thetable{B\arabic{table}}
\renewcommand\theequation{B\arabic{equation}}

\section{Fermion self-energy at finite temperature}
\label{sec:fermion self-energy}

The leading order contribution to the fermion self-energy is independent of momentum. This is because the momentum dependence carries an extra factor of $c$ \cite{PhysRevB.91.125136, PhysRevB.95.245109, PhysRevX.7.021010, PhysRevB.98.075140}. The temporal part of the fermion self-energy after scaling out $1/c$ is given by
\begin{equation}
\Sigma^{1L(1)}(k_0) = i \frac{3 \pi v}{2 c} T \sum_{\omega_n} \int \frac{d p_x \, dp_y}{(2 \pi)^2}
\frac{\omega_n + k_0}{\lt[ (\omega_n + k_0)^2 + (w p_x - p_y)^2 \rt] \lt[ |\omega_n| + |p_x| + c |p_y| + M \rt]},
\label{eq:fermion self energy finite T 1}
\end{equation}
where $\omega_n = 2 \pi T n$ with $n \in \mathbb{Z}$, and $k_0 = 2 \pi T (m + \frac12)$ with $m \in \mathbb{Z}$. Since we do not expect any logarithmic divergences in $v$, we can set $w = c = 0$ inside the integral, and do the $p_y$ and $p_x$ integrals
\begin{align}
\nn \Sigma^{1L(1)} &= i \frac{3 \pi v}{2 c} T \sum_{\omega_n} \int \frac{d p_x \, dp_y}{(2 \pi)^2}
\frac{\omega_n + k_0}{\lt[ (\omega_n + k_0)^2 + p_y^2 \rt] \lt[ |\omega_n| + |p_x| + M \rt]}
= i \frac{3 \pi^2 v}{2 c} T \sum_{\omega_n} \int \frac{d p_x }{(2 \pi)^2}
\frac{\sign(\omega_n + k_0)}{ |\omega_n| + |p_x| + M}
\\ &
= i \frac{3 \pi^2 v}{2 c} \frac{2}{(2 \pi)^2} T \sum_{\omega_n} 
\sign(\omega_n + k_0) \log\lt(\frac{|\omega_n| + M + \Lambda}{|\omega_n| + M}\rt),
\label{eq:fermion self energy finite T 2}
\end{align}
where $\L$ is the momentum cutoff. The infinite sum can be simplified
\begin{align} 
\Sigma^{1L(1)} =
i \frac{3}{4} w T \sign(k_0) \sum_{n = -\bar k_0}^{\bar k_0} \log\lt(1 + \frac{\Lambda}{|2 \pi T n| + M}\rt),
\label{eq:fermion self energy finite T 3}
\end{align}
where we have used the notation $\bar k_0 \equiv |\frac{k_0}{2\pi T}| - \f12$.
Up to now we could compute everything exactly. Now we take the limit of large momentum, $\L \gg \abs{k_0} + M$. We then have
\begin{align}
\nn 
\Sigma^{1L(1)} &\approx
i \frac{3}{4} w T \sign(k_0) \sum_{n = -\bar k_0}^{\bar k_0} \log\lt(\frac{\Lambda}{2 \pi T |n| + M} \rt)
= i \frac{3}{4} w T \sign(k_0) 
\lt(
\log \frac{\Lambda}{M}
+
2 \sum_{n = 1}^{\bar k_0} \log \frac{\Lambda}{2 \pi T n + M}
\rt)
\\ &=
i \frac{3}{4} w T \sign(k_0) 
\lt(
\log \frac{\Lambda}{M}
+
2 \log
\left(
\frac{\left(\frac{\Lambda }{2 \pi T}\right)^{\bar k_0} 
	\Gamma\left(1+\frac{M}{2 \pi T}\right)}
{\Gamma \left(1+\bar k_0+\frac{M}{2\pi T}\right)}
\right)
\rt).
\label{eq:fermion self energy finite T 4}
\end{align}
Since $\frac{M}{2\pi T} \rightarrow 0$ as a power of $v$ (c.f. Appendix \ref{sec:thermal mass}), we can neglect it to leading order,
\begin{eqnarray}
\nn 
\Sigma^{1L(1)} =
i \frac{3}{4} w T \sign(k_0) 
\lt(
\log \frac{\Lambda}{M}
+
2 \log
\left(
\frac{\left(\frac{\Lambda }{2 \pi T}\right)^{\bar k_0} }
{\Gamma \left(1+\bar k_0\right)}
\right)
\rt).
\end{eqnarray}
We convert back to the frequency $k_0$ and get
\begin{eqnarray}
\nn 
\Sigma^{1L(1)}(k_0,T) =
i \frac{3}{4} w \, T \sign(k_0) 
\lt(
2 \frac{|k_0| - \pi T }{2\pi T} \log
\left(\frac{\Lambda }{2\pi T}\right)
-
2 \log \lt[
\Gamma \left(1+\frac{|k_0| - \pi T}{2\pi T}\right)
\rt]
+ \log \frac{\Lambda}{M}
\rt).
\end{eqnarray}


\renewcommand\thefigure{C\arabic{figure}}
\renewcommand\thetable{C\arabic{table}}
\renewcommand\theequation{C\arabic{equation}}

\section{Numerical calculation of $\lambda_L^{(F)}$}
\label{sec:numerical Lyapunov exponent fermion}

Here we give some details on the numerical calculation of $f_F$ from Eq. (\ref{eq:matrix equation continuou f_F}). We convert the integral into a discrete sum and introduce a cutoff $\bar \L_0$ for the frequency summation. The equation becomes of the form $\mathcal{M} (\omega) f_F(\omega) = 0$, where $\mathcal{M}$ is a finite matrix given by 
\begin{equation}
\begin{split}
\mathcal{M}(\omega)_{k_0,k_0'}
& =
\frac{\Delta k_0}{2 \pi} \, 
i \frac{3}{4} w(v)
\lt( \pi - 2 \arctan\lt(\frac{\bar M}{|k_0 - k_0'|}\rt) \rt)
\lt( \sinh \frac{\abs{k_0 - k_0'}}{2} \rt)^{-1} 
- 
\delta_{k_0', k_0}
\\ &  
\times 
\Bigg[
\omega - i \frac{3}{2} w(v) 
\lt(
\lt(1 + \frac{i \omega}{2\pi}\rt)
\log\left(\frac{\bar \Lambda}{2\pi}\right)
+
\log \lt[
\Gamma \left(\f12 - \frac{i k_0}{2\pi}\right)
\rt]
+
\log \lt[
\Gamma \left(\f12 + \frac{i (k_0 - \omega)}{2\pi}\right)
\rt]
- \log\lt(\frac{\bar \Lambda}{\bar M}\rt)
\rt)
\Bigg],
\end{split}
\label{eq:eigenvalue problem lambda_F}
\end{equation}
where $k_0,k_0' \in [-\bar \Lambda_0, \bar \Lambda_0]$. We sweep values of $\omega$ on the positive imaginary axis, and plot the eigenvalue with the smallest magnitude, $\abs{E_0}$. The Lyapunov exponent is the largest value of $\lambda_L^{(F)} = -i \omega$ for which $\abs{E_0} = 0$. For a large enough $\bar \L_0$ the value of the integral doesn't change any more. We take $\bar \L_0 = 15$ and for each other parameter we decrease $\Delta k_0$ until the integral converges. We find that $\lambda_L^{(F)} \sim v \sim w^2$ up to logarithms. This means that the $O(w)$ terms in Eq. (\ref{eq:eigenvalue problem lambda_F}) cancel. The nonzero result is due to the thermal mass $M(T,\L,v)$. However, our calculation is not controlled up to $O(w^2)$, since we have not computed higher order rung corrections to the Bethe-Salpeter equation and higher order self-energy terms. Therefore, our result is null at the order up to which we have control.  


\renewcommand\thefigure{D\arabic{figure}}
\renewcommand\thetable{D\arabic{table}}
\renewcommand\theequation{D\arabic{equation}}

\section{Higher-order graphs for $f_F$}
\label{sec:higher-order graphs for f_F}

\begin{figure}[h]
	\subfigure[\, $\tilde K_F$]{\includegraphics[width=2.in]{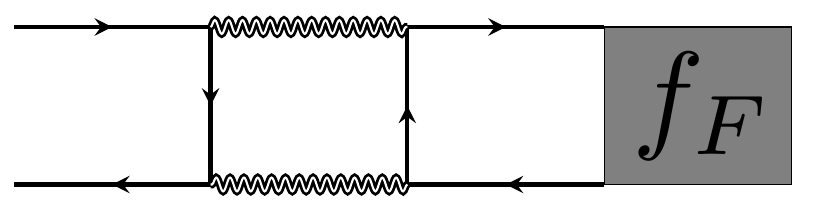}}
	\hspace{2mm}
	\subfigure[\, $\bar K_F$]{\includegraphics[width=2.in]{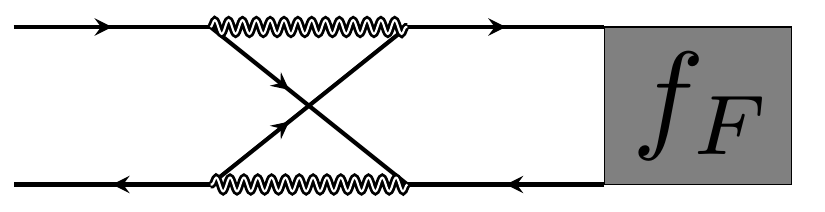}}
	\hspace{2mm}
	\subfigure[\, $\hat K_F$]{\includegraphics[width=2.in]{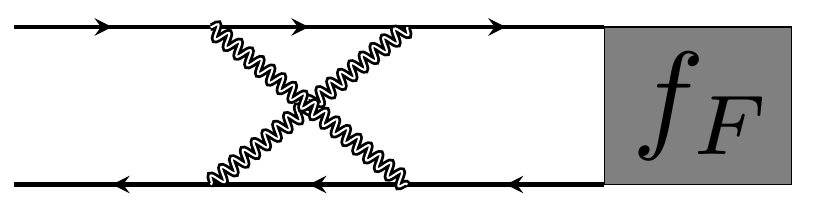}}
	\caption{
		The leading higher-order diagrams in the Bethe-Salpeter equation for $f_F$.
	}
	\label{fig:f_F higher order}
\end{figure}

Here we consider the higher order contributions to $f_F$ shown in Fig. \ref{fig:f_F higher order} and argue that they are at least of order $\O(w^2)$ up to logarithms.
The three kernels are given by
\begin{align}
\begin{split}
& \tilde K_F^{(n,m),(n',m')}(k, k', \omega) \sim  v^2 \int d^3 q 
\, D^R(q)
\, D^{A}(q - \omega) G^W_{n,\bar m} (k - q)
\, G^W_{n', \bar m'} (k' - q),
\end{split}
\\ 
\begin{split}
& \bar K_F^{(n,m)}(k, k', \omega) \sim  v^2 \int d^3 q 
\, D^R(q)
\, D^{A}(k + k' - q - \omega) G^W_{n,\bar m} (k - q)
\, G^W_{n, \bar m} (k' - q),
\end{split}
\\ 
\begin{split}
& \hat K_F^{(n,m)}(k, k', \omega) \sim  v^2 \int d^3 q 
\, G^R_{n,\bar m}(q)
\, G^A_{n,\bar m}(k + k' - q - \omega) D^W (k - q)
\, D^W (k' - q),
\end{split}
\label{eq:higher order kernels}
\end{align}
where $\tilde K_F$, $\bar K_F$ and $\hat K_F$ refer to diagrams (a), (b) and (c) in Fig. \ref{fig:f_F higher order}, respectively, and we have excluded numerical factors, since we are interested in the power of $v$ that the kernels scale with. Note that $\tilde K_F$ connects Green's functions at hot spot $(n,m)$ to $f_F^{(n',m')}$ at all others hot spots $(n',m')$, while the other two only connect the same hot spots. It is easy to check that once $1/c$ is scaled out of the $\vec q$ variables the integrals are UV finite (the exponential decay with large $q_0$ from the Wightman functions makes them even more UV safe than the equilibrium perturbative corrections). This implies the kernels come with a factor of $\sim v^2/c$. The kernels $\tilde K_F^{(n,m),(n',m')}$ that involve fermions from nearly perpendicular hot spots are finite even without scaling out $1/c$ and are therefore suppressed even further \cite{PhysRevX.7.021010, PhysRevB.95.245109}. After the $q$ integrations, the $\vec k'$ integration will give another factor of $1/c$, as in the lowest order rung diagram in Fig. \ref{fig:f_F Bethe-Salpeter}. Since all matrix elements of $\tilde{K}_F,\bar{K}_F,\hat{K}_F$ are then parametrically smaller than those of $K_F$, the resummation of the insertions in Fig. \ref{fig:f_F higher order} will only make a contribution to $\lambda_L^{(F)}$ of $\sim O(v^2/c^2) \sim O(w^2)$.


\renewcommand\thefigure{E\arabic{figure}}
\renewcommand\thetable{E\arabic{table}}
\renewcommand\theequation{E\arabic{equation}}

\section{Calculation of $K_B$}
\label{sec:calculation of K_B}

The kernel $K_B$ contains a fermion loop. Summing over the contributions from all the hot-spot pairs we get
\begin{align}
\nn
\begin{split}
& K_B(k, k', \omega) =  \sum_{n=1}^{4} \sum_{m = \pm} \lt(\frac{\pi v}{2}\rt)^2 \int \frac{d^3q}{(2\pi)^3} 
\, 
G^R_{(n,m)}(q) G^{A}_{(n,m)}(q - \omega)
\, G^W_{(n,-m)} (q - k)
\, G^W_{(n,-m)} (q - k')
\end{split}
\\ & \hspace{18mm} \equiv
\sum_{n=1}^{4} \sum_{m = \pm} K_B^{(n,m)}(k, k', \omega).
\label{eq:tilde K_1 1}
\end{align}
Here, the Green's functions and the Wightman functions are of the free fermions, since self-energy corrections would be higher order than the order we are working at. We start with $K_B^{(1,+)}(k, k', \omega)$,
\begin{equation}
\begin{split}
K_B^{(1,+)}(k, k', \omega)
& = \lt(\frac{\pi v}{2}\rt)^2 \frac{\pi^2}{v} 
\int \frac{d^3q}{(2\pi)^3} 
\, \frac{1}{q_0 + (q_x + q_y) + i \delta}
\, \frac{1}{q_0 - \omega + (q_x + q_y) - i \delta}
\\ & \, \frac{\delta((q_0 - k_0) + (q_x - v \, k_x) - (q_y - k_y))}{\cosh \frac{\beta (q_0 - k_0)}{2}}
\, \frac{\delta((q_0 - k_0') + (q_x - v \, k_x') - (q_y - k_y'))}{\cosh \frac{\beta (q_0 - k_0')}{2}}.
\end{split}
\label{eq:tilde K_1 2}
\end{equation}
We change variables to $q_+ = q_x + q_y, \, q_- = q_x - q_y$ and do the integration over $q_-$ by using the first delta function,
\begin{equation}
K_B^{(1,+)}(k, k', \omega) = 
\frac{\pi^3 v}{16} \int \frac{dq_0 \, dq_+}{(2\pi)^2} 
\, \frac{1}{q_0 + q_+ + i \delta}
\, \frac{1}{q_0 - \omega + q_+ - i \delta}
\, \frac{1}{\cosh \frac{\beta (k_0 - q_0)}{2}}
\, \frac{\delta((k'_0 - k_0) + v (k'_x - k_x) - (k'_y - k_y))}{\cosh \frac{\beta (k'_0 - q_0)}{2}}.
\label{eq:tilde K_1 compute 1}
\end{equation}
The integration over $q_+$ can be done via poles, and then the $q_0$ integration is trivial as well,
\begin{equation}
\begin{split}
K_B^{(1,+)}(k, k', \omega) &= \frac{\pi^3 v}{16} 
\, \frac{i \, \delta((k'_0 - k_0) + v (k'_x - k_x) - (k'_y - k_y))}{\omega}
\int \frac{dq_0}{2\pi}
\, \frac{1}{\cosh \frac{\beta (k_0 - q_0)}{2}}
\, \frac{1}{\cosh \frac{\beta (k'_0 - q_0)}{2}}
\\ &= 
\frac{i}{\omega}
\, \frac{\pi^2 v}{16} 
\, \frac{k_0 - k_0'}{\sinh \frac{\beta (k_0 - k_0')}{2}}
\, \delta((k'_0 - k_0) + v (k'_x - k_x) - (k'_y - k_y)).
\end{split}
\label{eq:tilde K_1 compute 2}
\end{equation}
The result is trivially extended to all other $K_B^{(n,m)}(k, k', \omega)$, and we can sum all contributions to get
\begin{equation}
K_B(k, k', \omega)
= \frac{i}{\omega}
\, \frac{\pi^2 v}{16} 
\, \frac{k_0 - k_0'}{\sinh \frac{\beta (k_0 - k_0')}{2}} 
\sum_{n=1}^{4} \sum_{m = \pm}	
\delta((k'_0 - k_0) + e_n^{(m)}(\vec{k}' - \vec k)).
\label{eq:tilde K_1 answer}
\end{equation}


\renewcommand\thefigure{F\arabic{figure}}
\renewcommand\thetable{F\arabic{table}}
\renewcommand\theequation{F\arabic{equation}}

\section{Numerical calculation of $\lambda_L^{(B)}$}
\label{sec:calculation of lambda_L^B}

\begin{figure}[!ht]
	\subfigure[]
	{\includegraphics[width=2.5in]{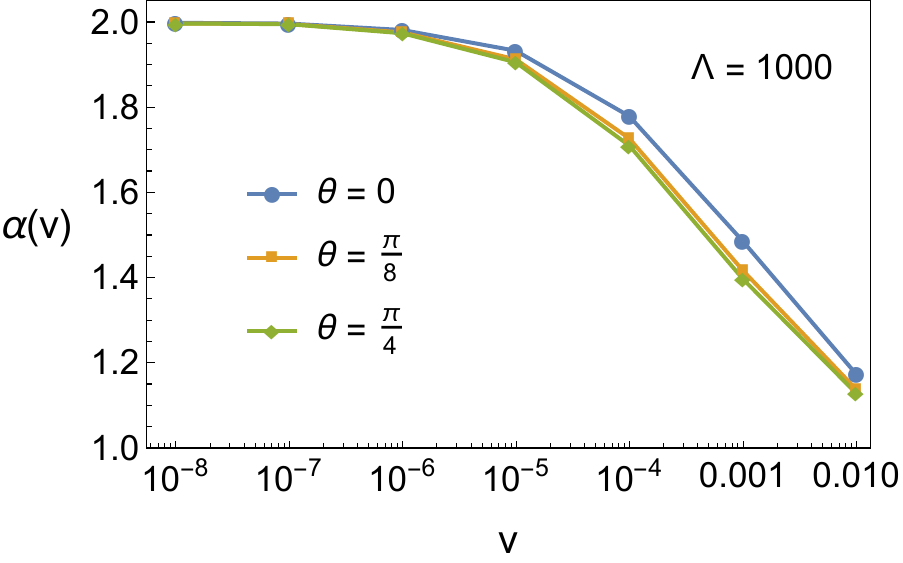}
	    } ~~~~~
	\subfigure[]
	{\includegraphics[width=2.5in]{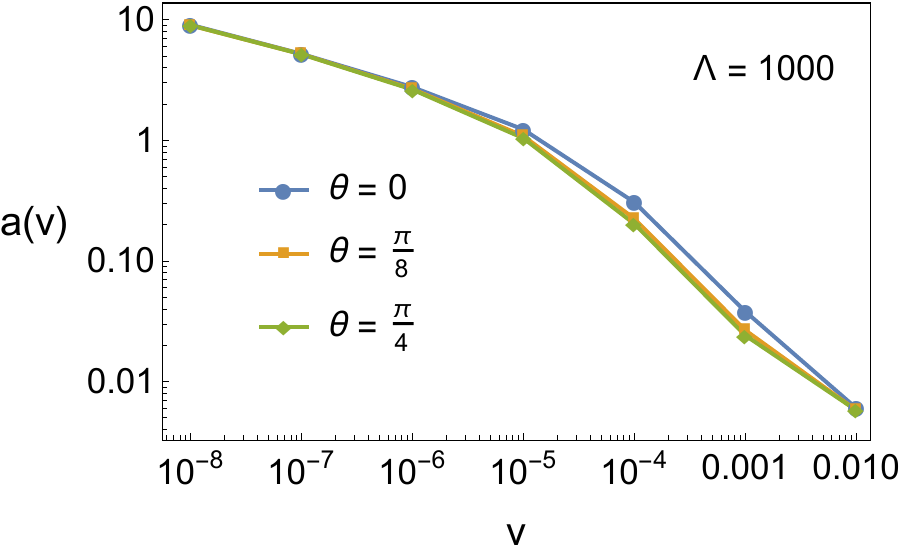}
	    } ~~~~~
	\caption{The form of $\alpha(v,10^3,\theta)$ and $\a(v,10^3,\theta)$ as functions of $v$ for various $\theta$. The dependence is weak: $\alpha$ changes by less than $5\%$ and $a$ changes by a factor of less than $1.5$. We note that because of the $C_4$ symmetry in the problem we can focus only on $\theta \in (0,\pi/4)$.}
	\label{fig:finite p plots changing theta}
\end{figure}

Here we provide some more details about the numerical calculation of $\lambda_L^{(B)}$. We first note the transition from $\mathcal{M}_{3\text{D}}$ to $\mathcal{M}_{2\text{D}}$ is additionally justified by the fact that for a fixed $\Delta k_x, \Delta k_y$ the largest $\lambda_L^{(B)}$ computed from $\mathcal{M}_{2\text{D}}$ is always larger than that of $\mathcal{M}_{3\text{D}}$. Since we are ultimately interested the $k$-integral of the fastest growing eigenvector, $\int dk f(i \lambda_L^{(B)}, k)$, this implies that the ansatz we make is safe. 
\\
\indent 
The ratio of momentum cutoff over temperature $\bar \L = \L/T$ appears in two different places in Eq. (\ref{eq:M 2D}): as the largest momentum index, and as a parameter in $\bar M(\bar \Lambda,v)$. The actual physical $\bar \Lambda$ can be quite large, but we don't have the computational power to go to the necessary resolution except for modest $\bar \Lambda$. However, the dependence of the integral (or sum) on the cutoff is weak, i.e. the result is a series in the inverse cutoff, since the integral is convergent (we confirm this). On the other hand, the dependence on $\bar \Lambda$ from $\bar M(\bar \Lambda,v)$ is logarithmic.
Therefore, in order to find the Lyapunov exponent for large values of $\bar \Lambda$, we can treat the integration cutoff and $\bar \Lambda$ as separate. For each given $10 \leq \bar \Lambda \leq 3 \times 10^5$, we find the Lyapunov exponent for several (mostly) much smaller cutoffs and then extrapolate the answer to the infinite cutoff limit. Our exact extrapolation procedure is as follows: we first compute $\lambda_L^{(B)}$ for some fixed cutoff and momentum spacing $\Delta k$, and from Eq. (\ref{eq:Lyapunov exponent boson}) we extract the fit for $h(\bar \Lambda)$ of the form in Eq. (\ref{eq: h fit}). Then holding the cutoff fixed we recompute the numerical coefficients of $h(\bar \Lambda)$ for several decreasing values of $\Delta k$, and extrapolate these results to the $\Delta k \rightarrow 0$ limit using a linear fit. We then assume that the slope of that fit is independent of $\bar \Lambda$, which holds for the values we have checked, and perform the infinite cutoff extrapolation on the $\Delta k \rightarrow 0$ values of the numerical coefficients in $h(\bar \Lambda)$, giving us the values in Eq. (\ref{eq: h fit}).
\\
\indent 
For a finite external momentum $\vec p$, Eq. (\ref{eq:f_B Bethe Salpeter calculation 2}) becomes (before setting $v=0$ in the dispersions)
\begin{align}
\nn & 
\int d^3k'
\Bigg[
i \, \frac{3 \, v}{2^6 \pi}
\frac{k_0 - k_0'}{\sinh \frac{\beta (k_0 - k_0')}{2}} 
\sum_{n=1}^{4} \sum_{m = \pm}	
\frac{\delta\lt( (k'_0 - k_0) + e_n^{(\bar m)}(\vec{k}' - \vec k) \rt)}{\omega + e_n^{(m)}(\vec p)} 
\\ & -
\delta(k' - k)
\[- i k_0 + c(v) \( |k_x| + |k_y| \) + M \] \[ i (k_0 - \omega) + c(v) \( |k_x - p_x| + |k_y - p_y| \) + M \] 
\Bigg]
f_B(\omega, k') = 0.
\label{eq:matrix equation f_B Bethe Salpeter external p}
\end{align}
As before, we set $v = 0$ in the dispersions and assume an ansatz of the form $f_B(\omega, \vec p, k) = g_B(\omega, \vec p, \vec k) \,  \delta(k_0)$. The resulting two dimensional equation is given in Eq. (\ref{eq:M 2D external p}). Solving it gives the deviation from the $\vec p = 0$ Lyapunov exponent in \eq{eq:Lyapunov exponent small p}. In Fig. \ref{fig:finite p plots changing theta} we include some plots of the weak $\theta$ dependence of $a(v,\Lambda, \theta)$ and $\a(v,\Lambda, \theta)$.

\end{document}